\documentclass[preprint]{aastex}

\begin{document}

\title{VARIABILITY OF DISK EMISSION IN PRE-MAIN SEQUENCE AND RELATED STARS. III. Exploring Structural Changes \\in the Pre-transitional Disk in HD 169142.}
\shorttitle{VARIABILITY OF DISK EMISSION. III.}
\shortauthors{Wagner, Sitko, Grady et al.}

\author{ Kevin R. Wagner,\altaffilmark{1} \altaffilmark{2}  \altaffilmark{3} Michael L. Sitko,\altaffilmark{1} \altaffilmark{2} \altaffilmark{3} Carol A. Grady,\altaffilmark{4} \altaffilmark{5} \altaffilmark{6} Barbara A. Whitney,\altaffilmark{2} \altaffilmark{7} Jeremy R. Swearingen,\altaffilmark{1} Elizabeth H. Champney,\altaffilmark{1} \altaffilmark{3} Alexa N. Johnson,\altaffilmark{1} Chelsea Werren,\altaffilmark{1} \\Ray W. Russell,\altaffilmark{3} \altaffilmark{8} Glenn H. Schneider,\altaffilmark{9} Munetake Momose,\altaffilmark{10} Takayuki Muto,\altaffilmark{11} Akio K. Inoue,\altaffilmark{12} James T. Lauroesch,\altaffilmark{13} Alexander Brown,\altaffilmark{14} Misato Fukagawa,\altaffilmark{15}  Thayne M. Currie,\altaffilmark{16} Jeremy Hornbeck,\altaffilmark{13} John P. Wisniewski,\altaffilmark{17} and Bruce E. Woodgate\altaffilmark{18}}

\altaffiltext{1}{Department of Physics, University of Cincinnati, Cincinnati, OH 45221-0011}
\altaffiltext{2}{Space Science Institute, 4750 Walnut St., Suite 205, Boulder, CO 80301}
\altaffiltext{3}{Visiting Astronomer at the Infrared Telescope Facility, which is operated by the University of Hawaii under Cooperative Agreement no. NNX-08AE38A with the National Aeronautics and Space Administration, Science Mission Directorate, Planetary Astronomy Program.}
\altaffiltext{4}{Eureka Scientific, 2452 Delmer, Suite 100, Oakland, CA 96002}
\altaffiltext{5}{ExoPlanets and Stellar Astrophysics Laboratory, Code 667, NASA's
Goddard Space Flight Center, Greenbelt, MD 20771}
\altaffiltext{6}{Goddard Center for Astrobiology}
\altaffiltext{7}{Department of Astronomy, University of Wisconsin, 475 N. Charter St., Madison, WI 53706-1582}
\altaffiltext{8}{The Aerospace Corporation, Los Angeles, CA 90009}
\altaffiltext{9}{Steward Observatory, 933 N. Cherry Avenue, University of Arizona, Tucson, AZ 85721}
\altaffiltext{10}{Ibaraki University, Japan, 〒310-0056 Ibaraki, Mito, Bunkyo, ２丁目1−1}
\altaffiltext{11}{Kogakuin University, 1-24-2 Nishishinjuku, Shinjuku, Tokyo 163-8677, Japan}
\altaffiltext{12}{Osaka Sangyo University, College of General Education, 3-1-1 Nakagaito, Daito, Osaka 574-8530, Japan}
\altaffiltext{13}{University of Louisville Research Foundation, Inc., 2301 S 3rd St, Louisville, KY 40292}
\altaffiltext{14}{Center for Astrophysics and Space Astronomy, Astrophysics Research Laboratory, 593 UCB, University of Colorado, Boulder, CO 80309-0593}
\altaffiltext{15}{Department of Earth and Space Science, Graduate School of Science, Osaka University, 1-1, Machikaneyama, Toyonaka, Osaka 560-0043, Japan}
\altaffiltext{16}{Oak Ridge Associated Universities, 100 ORAU Way, Oak Ridge, TN 37830-6218}
\altaffiltext{17}{University of Oklahoma, 660 Parrington Oval, Norman, OK 73019}
\altaffiltext{18}{NASA Goddard Space FLight Center, 8800 Greenbelt Rd, Greenbelt, MD 20771}

\begin{abstract}

We present near-IR and far-UV observations of the pre-transitional (gapped) disk in HD 169142 using NASA's Infrared Telescope Facility and Hubble Space Telescope. The combination of our data along with existing data sets into the broadband spectral energy distribution reveals variability of up to 45\% between $\sim$1.5-10 $\mu$m over a maximum timescale of 10 years. All observations known to us separate into two distinct states corresponding to a high near-IR state in the pre-2000 epoch and a low state in the post-2000 epoch, indicating activity within the $\lesssim$1 AU region of the disk. Through analysis of the Pa $\beta$ and Br $\gamma$ lines in our data we derive a mass accretion rate in May 2013 of \.{M} $\approx$ (1.5 - 2.7) x 10$^{-9}$ M$_{\odot}$ yr$^{-1}$.  We present a theoretical modeling analysis of the disk in HD 169142 using Monte-Carlo radiative transfer simulation software to explore the conditions and perhaps signs of planetary formation in our collection of 24 years of observations.  We find that shifting the outer edge ($r\approx 0.3$ AU) of the inner disk by 0.05 AU toward the star (in simulation of accretion and/or sculpting by forming planets) successfully reproduces the shift in NIR flux.  We establish that the $\sim$40-70 AU dark ring imaged in the NIR by \cite{quanz13} and \cite{momose13} and at 7 mm by \cite{osorio14} may be reproduced with a 30\% scaled density profile throughout the region, strengthening the link to this structure being dynamically cleared by one or more planetary mass bodies. 

\end{abstract}

\keywords{Stars: individual (HD 169142) -- stars: variables (Herbig Ae/Be) -- planetary systems: protoplanetary disks -- planetary systems: planet-disk interactions}

\section{Introduction}

To study the regions of planetary formation within disks of gas and dust around young stars, we look toward the stars around us in the galaxy to identify those which may offer snapshots of these systems in their various stages of evolution. As these systems age, multiple mechanisms begin to clear their primordial disks of material. Because planets may be forming from this material, the details of how and when these disks become cleared are important to understand the origins and evolution of planetary systems. Some of these mechanisms for clearing are accretion onto the central star, photo-evaporation, grain growth and coagulation into larger bodies, and large scale sculpting by orbiting companions. In the absence of strong external illumination by hot nearby stars, photo-evaporation may only be responsible for clearing material from the inner regions outward in the disk, and is counteracted by the viscous accretion onto the central star. The other two mechanisms may operate at any radius within the disk.\footnotemark \footnotetext{See \cite{williams11} for a detailed discussion on various disk clearing mechanisms.} \cite{strom89} was first to suggest that we should see disks with cleared inner regions -- in transition to becoming completely cleared, and leading to the class designation of ``transitional" disks. These disks have since been confirmed to exist \citep{espaillat10}, and some objects also display evidence in their spectra for the presence of structures inside of these cleared regions, ruling out photo-evaporation as a dominant mechanism for disk clearing. These objects are designated as ``pre-transitional disks" and are thought to be in earlier stages of evolution compared to the transitional disks \citep{espaillat07}. Given their long orbital periods, changes to structures in the outer disk proceed over timescales which may be longer than the lifetime of any single human being. However, material inside of the gap at small radii may orbit on periods short enough to allow for changes to propagate through the disk and become evident in the spectral energy distribution (hereafter SED) over timescales shorter than a decade (see papers I and II in this series: \cite{sitko08} and \cite{sitko12}, along with \cite{espaillat11}, \cite{muzerolle09}, and Flaherty et al. (2011 and 2012)).\footnotemark \footnotetext{Notably, \cite{espaillat11} detect variability in 12 out of 14 T Tauri stars observed with Spitzer/IRS over periods of 2-3 years.} Hence there may not exist one single model which adequately describes some objects, while multiple models of these objects in their different states allow for speculation on the conditions and mechanisms responsible for producing the variable emission, and perhaps insight on the evolution of planet forming regions at approximately terrestrial orbital radii. 

Additionally, high-contrast imagery has been performed on a number of young stars hosting protoplanetary disks. These images show structures of disks in various stages of evolution such as rings, gaps, shadows, spirals, asymmetric bright/dark regions, and other features which may be indicative of conditions and perhaps companions within the disk. However, it is not always the case that the structures responsible for producing these features are able to be identified through the images alone. In particular, dark bands which appear to be annular gaps may in fact be shadows cast by inner disk structures. To determine the true nature of these features, Monte-Carlo radiative transfer simulation may first be used to determine the extent of shadows cast by structures which are well constrained by the SED. Once these structures (and their shadows) are determined, it is possible to explore the density and scale height conditions which reproduce the gap-like features of the images. Determining which stellar systems possess gaps in their disks may help to select targets for future studies of young planetary systems, and determining the locations and compositions of these gaps may provide direct insight about the conditions of immediate regions of planet formation. 

HD 169142 is a 6$^{+6}_{-3}$ Myr old Herbig Ae star at a distance of 145 pc which attracts particular interest since it has been identified as hosting a disk with an inner gap comparable to the size of our own solar system (\cite{meeus10}, \cite{honda12}). The disk is in the pre-transitional stage, characterized by emission from warm material close to the central star ($<$ 1 AU) as well as emission from cooler material further away from the star, with the inner gap residing between the two rings (\cite{honda12}, \cite{maaskant14}, \cite{osorio14}). The main disk extends to $\sim$250 AU from the central star (\cite{grady07}, \cite{quanz13}) and is seen at $\sim$13$^{\circ}$ from face-on, making it a good target for imagery of finer structures.  Near infrared (NIR) polarized light images of HD 169142 were obtained in 2011 by \cite{momose13} using Subaru/HiCIAO and in 2013 by \cite{quanz13} using VLT/NACO, showing features of the disk that were previously unresolved.\footnotemark \footnotetext{These images are reproduced later in this article in Figure 10.} The main structures revealed in these images are depicted in Figure 1.

\begin{figure}[htpb]
\figurenum{1}
\epsscale{.45}
\plotone{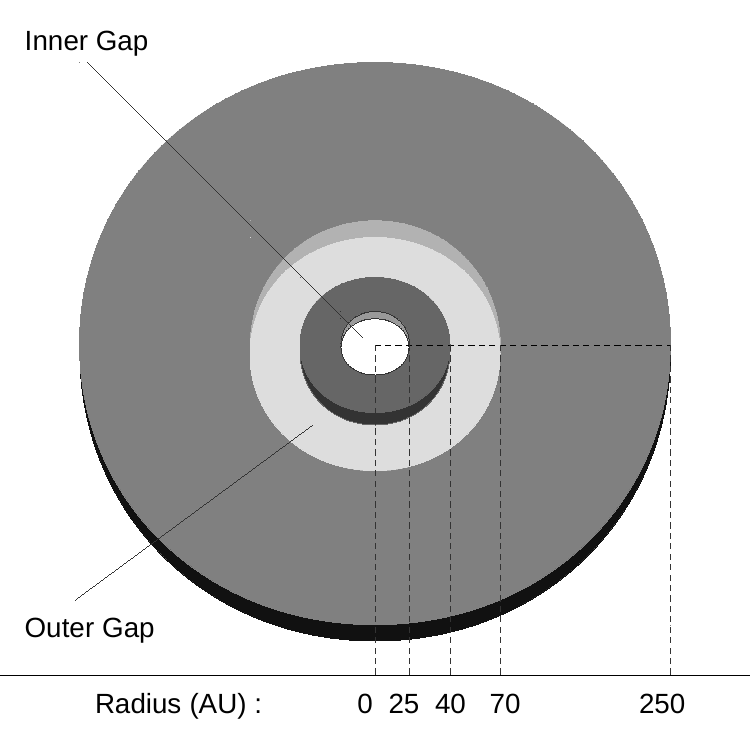}
\caption{\footnotesize The main structures of the pre-transitional disk in HD 169142 revealed through high contrast and angular resolution imagery. Note that the sub-AU components (the central star and the inner disk) are not visible on this scale.}
\end{figure}

 In these images, the inner gap is clearly resolved to a radius of $\sim$25 AU and appears to be devoid of scattering material, with most of the scattered light emanating from the inner edge of the outer disk. Also visible in the images is a dark annular ring, reportedly located between $\sim$40-70 AU  of the central star by Quanz et al. and Osorio et al., and between $\sim$50-80 AU by Momose et al. The interior of this region does not appear to be entirely cleared of scattering material (unlike the gap inside r=25 AU), as each team detects emission above instrumental detection levels from within the ring. The feature is confirmed at 7-mm \citep{osorio14}, indicating that it is at least in part due to a deficit in density extending through the disk mid-plane (i.e. a second gap).

We present an investigation of the HD 169142 system through computer modeling using Hochunk3D, a Monte-Carlo radiative transfer simulation software -- described in \cite{whitney13}. Since not all structures are well constrained by the SED alone, to understand these structures within the disk with greater accuracy we pair SED modeling with modeling to fit the observed properties of the multi-band imagery. We combine modeling of the multi-epoch SED with single epoch multi-band imagery to see what we may learn about the disk and environment of potential planet forming regions in HD 169142.

\section{Observations and Data Reduction}

\subsection{SpeX Observations}

HD 169142 was observed using the SpeX spectrograph \citep{rayner03} on NASA's Infrared Telescope Facility (IRTF). The SpeX observations were obtained  using the cross-dispersed (hereafter XD) echelle gratings in both short-wavelength mode (SXD) covering 0.8-2.4 \micron{} and long-wavelength mode (LXD) covering 2.3-5.4 \micron{}.  The LXD data were obtained on 2013 May 15 (UT), and the SXD on 2013 May 16 (UT). For both sets of data,  a 0.8 arcsec wide slit was used. The spectra were corrected for telluric extinction, and flux calibrated against the A0V star HD 162220  using the Spextool software \citep{vacca03,cushing04} running under IDL. The LXD spectrum was merged to the SXD, and required no re-scaling in the wavelength region where they overlapped.  

Due to the light loss introduced by the 0.8 arcsec slit used to obtain the XD spectra, changes in telescope tracking and seeing between the observations of HD 169142  and a calibration star may result in merged XD spectrum with a net zero-point shift compared to their true absolute flux values. To correct for any zero-point shift that results from the 0.8 arc sec echelle observations, we also observed HD 169142 on 2013 May 16 (UT) with the low dispersion prism in SpeX using a 3.0 arcsec wide slit. On nights where the seeing is 1.0 arc sec or better, past experience shows that this technique yields absolute fluxes that agree with aperture photometry to within 5\% or better. At the time of the observations, the seeing was $\sim$0.7 arc sec. For the prism observations, HD 162220 was used as the flux standard. The resulting scaling factor derived for the echelle spectrum was 0.95 (shift to fainter absolute fluxes by 5\%), which has been applied to the data.

\subsection{BASS Spectrophotometry}

We observed HD 169142 with The Aerospace Corporation's Broad-band Array Spectrograph System (BASS) on the IRTF on 2008 September 4 (UT) and 2011 August 1 (UT). BASS uses a fixed circular entrance aperture with a diameter of 1 mm. Gradual improvements in the internal optics have increased the effective projected diameter on the sky using the IRTF from ~3.4 arcsec in 2008 to ~4.4 arcsec in 2011. The light then passes through a cold beamsplitter to separate the light into two separate wavelength regimes. The short-wavelength beam includes light from 2.9-6 $\mu$m, while the long-wavelength beam covers 6-13.5 $\mu$m. Each beam is dispersed onto a 58-element Blocked Impurity Band (BIB) linear array, thus allowing for simultaneous coverage of the spectrum from 2.9-13.5 $\mu$m. The spectral resolution $R = \lambda$/$\Delta\lambda$ is wavelength-dependent, ranging from about 30 to 125 over each of the two wavelength regions \citep{hackwell90}. $\alpha$ Boo was used as an absolute flux calibration star for both observations \citep{russell12}. 

\subsection{HST/NICMOS Photometry}

Photometric observations of HD 169142 were obtained on 2005 April 30 (UT) using the Near Infrared Camera and Multi-Object Spectrometer (NICMOS) instrument on the Hubble Space Telescope. Images were obtained using the F110W in  and F171M filters. For the F171M data, the observations were obtained during target acquisition (ACQ mode). The F110W data were obtained after coronagraphic observations by slewing the target from behind the coronagraphic spot, and observing them in the direct (MULTIACCUM) mode, using an 8-point target dither. The F110W were then extracted using a point spread function for an A1 spectral type star. We obtain fluxes of F($\lambda$=1.12 $\mu$m)=5.567$\times 10^{-12}$ W $m^{-2}$ and F($\lambda$=1.60 $\mu$m)=2.836$\times 10^{-12}$ W $m^{-2}$, each with an uncertainty of 1\%. 

\subsection{HST/STIS and HST/COS Observations}

HD 169142 was observed 3 times in the FUV in 2013 using the Hubble Space Telescope: twice by the Space Telescope Imaging Spectrograph (STIS, see Woodgate et al. 1998), and once using the Cosmic Origins Spectrograph (COS, see Green et al. 2012). Table 1 is the journal of FUV observations, augmented by archival data for Altair using the Goddard High-Resolution Spectrograph (GHRS), which serves as a useful comparison star in the UV. All spectra were processed using the pipeline software appropriate for their date of observation (COS: Holland et al. 2014; STIS: Hernandez, S. et al. 2014; and HRS: Soderblom et al. 1995). 

\begin{table}[ht]
\centering
\begin{tabular}{cccccccc}
\multicolumn{8}{c}{Table 1: Journal of Hubble Space Telescope FUV Spectra}\\
\hline
\hline
Star & Date & PID & Obs. ID & Inst./Mode & Aper. & Cenwave PA & t$_{exp} (s)$\\
\hline
HD 169142 & 2013-04-05 & 13032 & LC0M02010  & COS/G130M & PSA & 1290 & 400\\
" & " & " & LC0M02020 & " & " & 1309 & 400\\
" & " & " & LC0M02030 & " & " & 1327 & 760.032\\
\hline
" & 2013-07-18 & 13032 & OC0M03010 & STIS/G140M & 52x0.2 & 1218 & 2040\\
" & " & " & OC0M03020 & " & " & 1218 & 2750\\
" & 2013-07-11 & 12996 & OC2H02020 & STIS/E140M & 0.2x0.2 & 1425 & 1732.1\\
" & " & " & OC2H02030 & " & " & 1425 & 2761.182\\
\hline
Altair & 1996-08-28 & 6446 & Z3GB0104T & GHRS/G140L & 2.0 & 1425 & 217.6\\
\hline
\end{tabular}
\end{table}

\subsection{SED and Variability}

\begin{figure}[htpb]
\figurenum{2}
\epsscale{.75}
\plotone{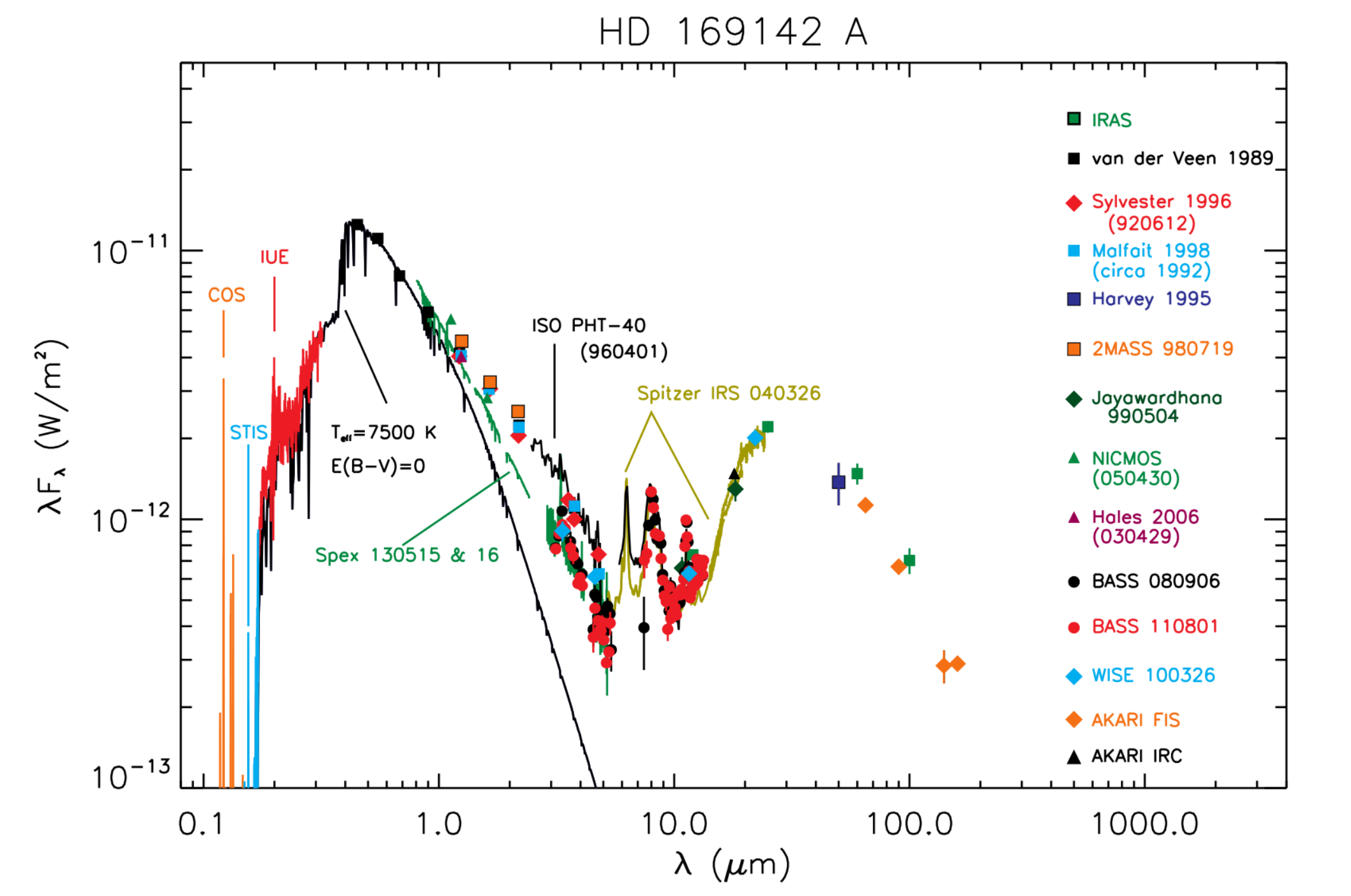}
\caption{\footnotesize HD 169142 spectral energy distribution (SED). At short wavelengths, the solid black line represents the unreddened photospheric spectrum. Note that  author names typically refer to data from publications, while instrument names refer to archival or previously unpublished data. }
\end{figure}

\begin{figure}[htpb]
\figurenum{3}
\epsscale{.75}
\plotone{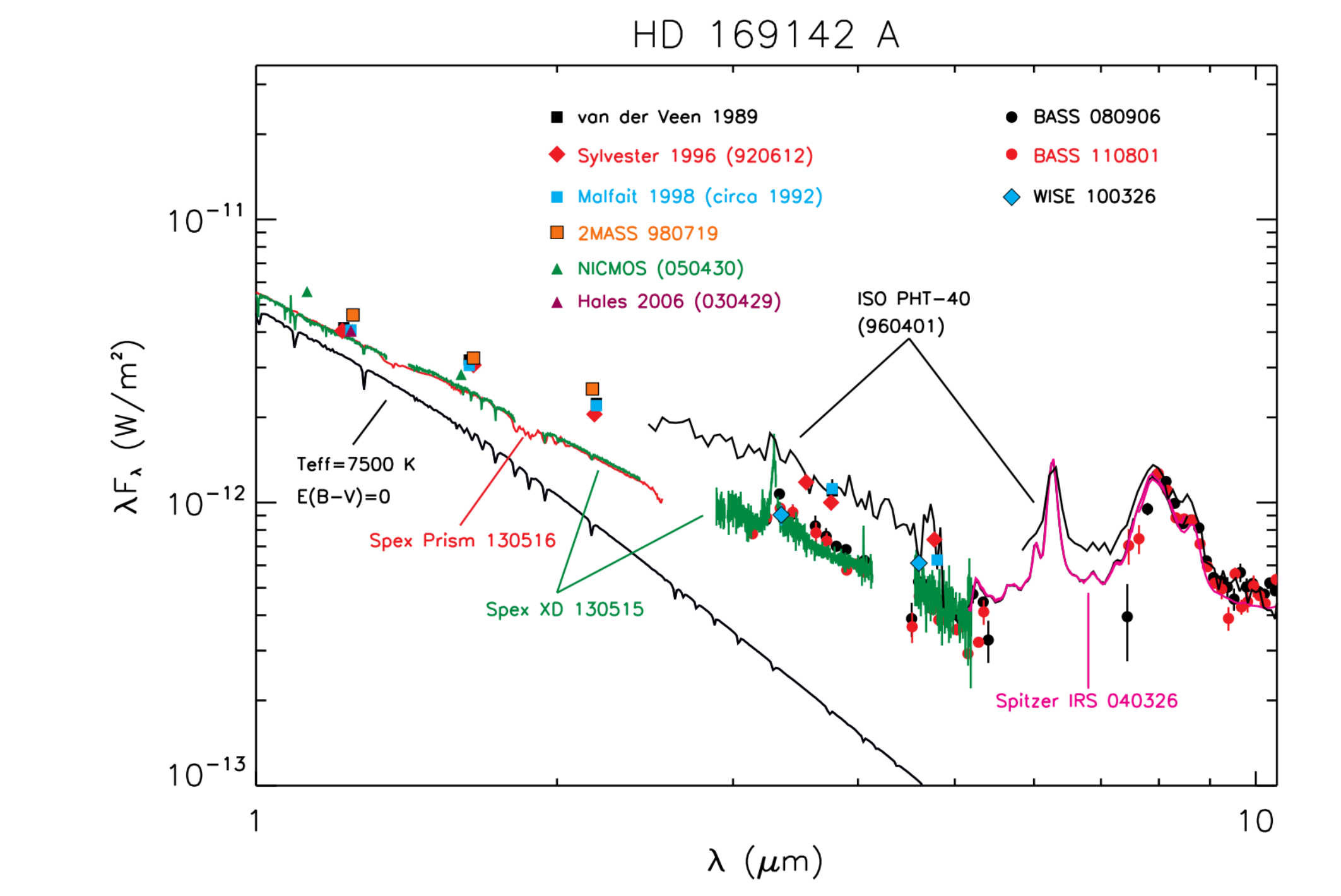}
\caption{\footnotesize Closeup of the NIR SED of HD 169142 showing the high (pre-2000) and low (post-2000) states. Note that the poly-cyclic aromatic hydrocarbon (PAH) emission bands do not participate in the variability.  }
\end{figure}

In addition to the data presented above, we have collected a multitude of archival and published data to aid in our construction of the broadband multi-epoch SED for HD 169142 (Figure 2). Our construction of the SED reveals variability in the NIR spectrum, specifically at wavelengths between 1.5-10 microns, which is confirmed through multiple observations. Measurements fall into two distinct flux states, with those recorded after the year 2000 exhibiting up to 45\% lower NIR flux than those recorded prior to 2000, with the change occurring over no more than 10 years -- as given by the 2MASS data (1998 July 19) and the IRTF/BASS data (2008 September 6). A close-up of the variable near-IR spectrum is presented in Figure 3. In $\S$3.5 we present an investigation on the nature of this variability. 

Data from the infrared spectrograph (IRS) on the Spitzer Space Telescope were obtained from the Spitzer Heritage Archive. ISO PHT-40 data were obtained from the Infrared Space Observatory (ISO) archives \citep{acke04}. Ultraviolet spectroscopy was obtained from the IUE archives. Our complete collection of photometric data can be found in Appendix B. What we identify as the post-2000 SED in this study consists of photometry from HST/NICMOS (2005), \cite{hales06}, the AKARI/IRC Point Source Catalog, and the AKARI/FIS Bright Source Catalog, along with spectroscopy from HST/COS, IRTF/BASS, IRTF/SpeX, and Spitzer/IRS. The pre-2000 SED consists of photometry from \cite{vanderveen89}, \cite{harvey95}, \cite{sylvester96}, \cite{malfait98}, \cite{jaya01} (observations from 1999), the 2MASS All Sky Point Source Catalog, and the IRAS Point Source Catalog, along with IUE and ISO PHT-40 spectroscopy.

\section{Results}

\subsection{Mass Accretion Rate}

Despite the presence of gaps in the inner regions of HD 169142, it is still likely experiencing a modest amount of gas accretion onto the star. Using spectra obtained with the \textit{International Ultraviolet Explorer}, \citet{grady07} derived a mass accretion rate in the 1990s of 

\begin{center}
\.{M} $\leq$ (0.7 - 1.8) x 10$^{-9}$ M$_{\odot}$ yr$^{-1}$.
\end{center}

The mass accretion rate at the time of the  SpeX data from 15-16 May 2013 (UT) can be derived using the net emission from the Pa $\beta$ and Br $\gamma$ lines. In order to do so, we modeled the underlying continuum in a manner described by \citet{sitko12}. Here the underlying emission of the stellar photosphere and dust emission were modeled using a SpeX spectrum of HD 98058 plus a modified blackbody to represent the dust emission, which were subtracted from the observed spectrum of HD 169142. The results are shown in Figure 4.  Using a distance of 145 pc, $M_{star}=1.65 M_{\odot}$, and $R_{star}=1.75 R_{\odot}$, we derive a mass accretion rate of 

\begin{center}
\.{M} $\approx$ (1.5 - 2.7) x 10$^{-9}$ M$_{\odot}$ yr$^{-1}$.
\end{center}

For the conversion of line luminosity to mass accretion rate, we used the calibrations of \citet{calvet04} and  \citet{muzerolle98} for Pa $\beta$ and Br $\gamma$, respectively.

\begin{figure}[h]
\figurenum{4}
\epsscale{1.1}
\plottwo{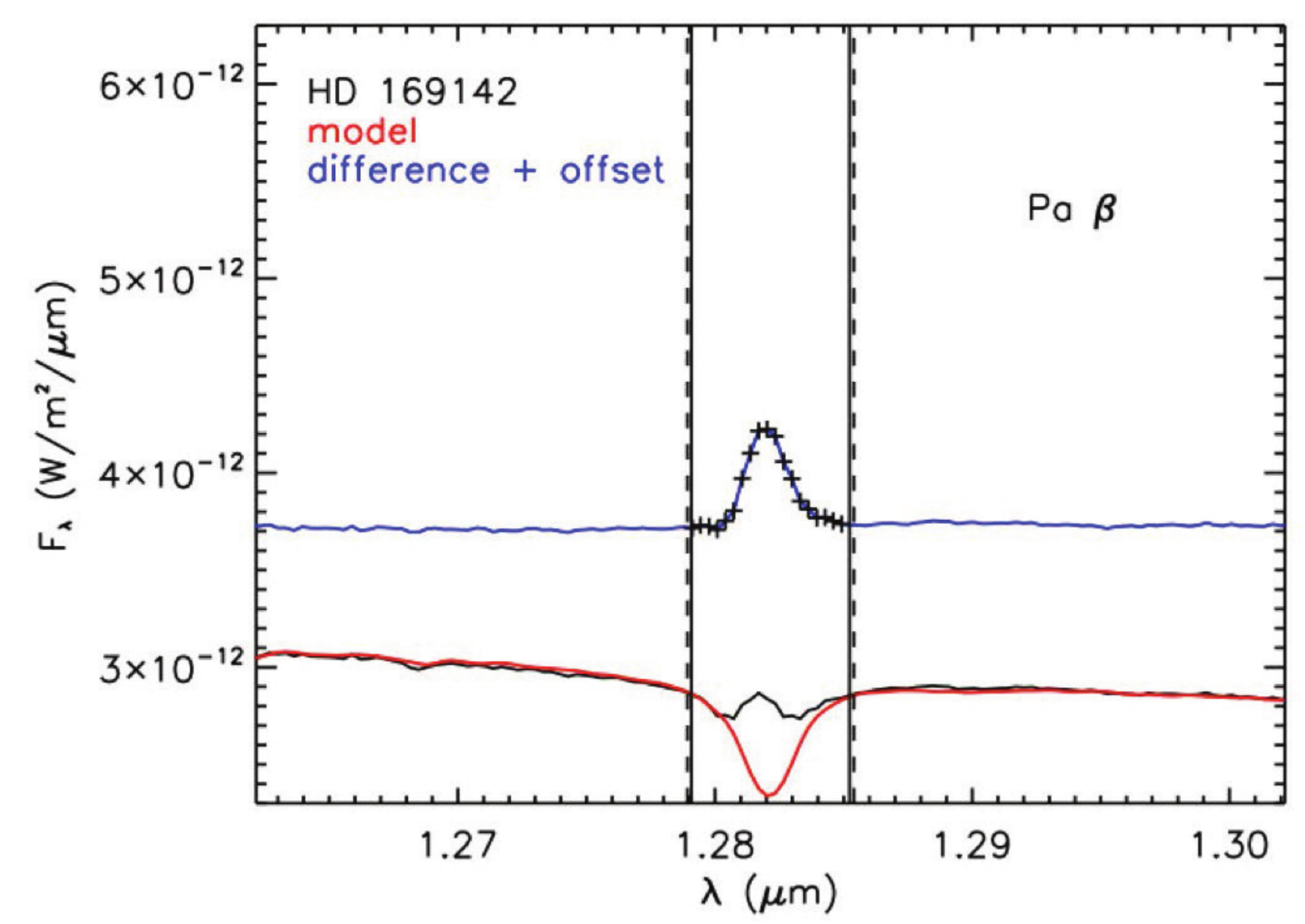}{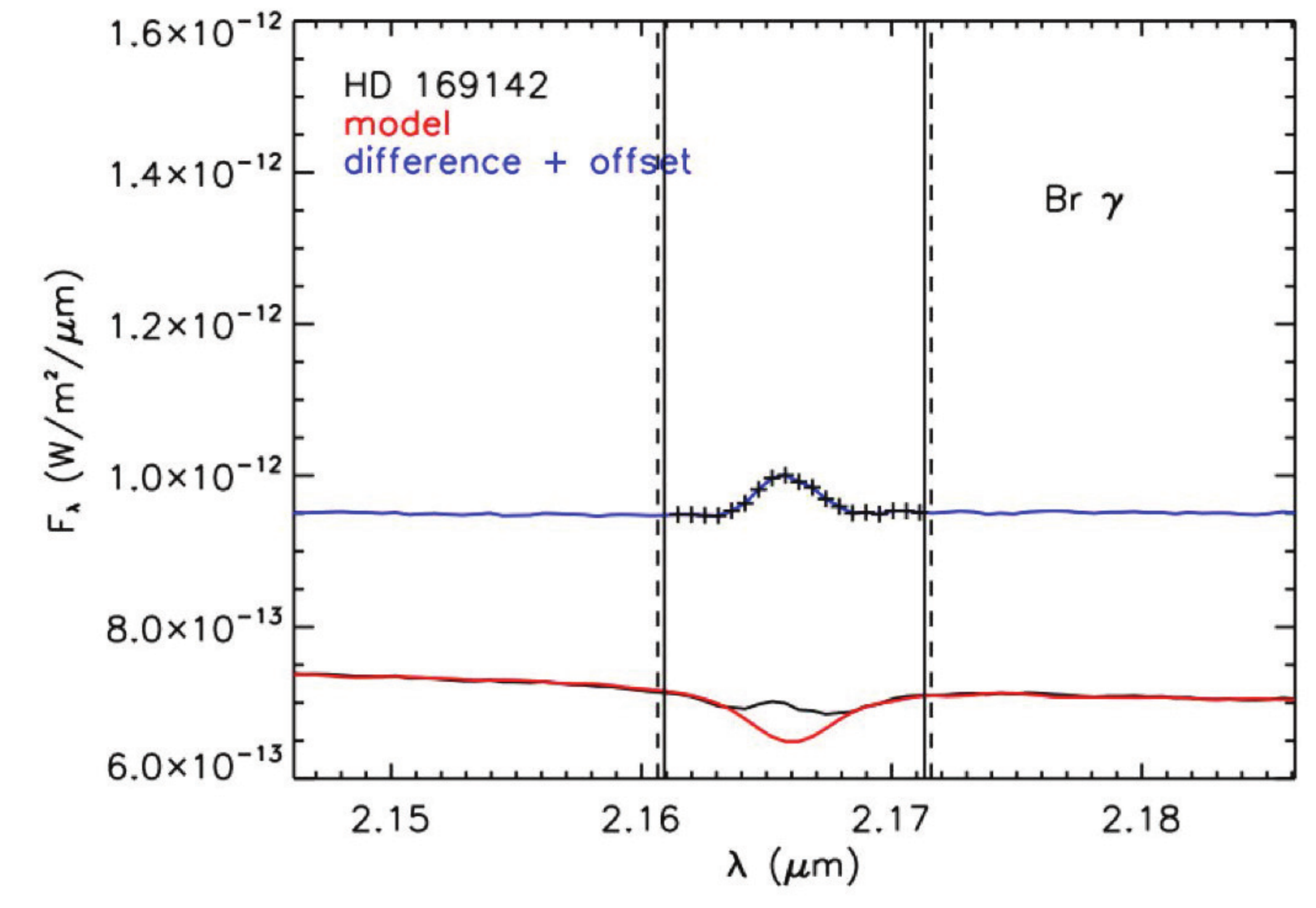}
\caption{\footnotesize The extracted emission line intensities of the Pa $\beta$ (left) and Br $\gamma$ (right) lines. In order to determine the strengths of the lines, we modeled the underlying background as a combination of an A7 IVn SpeX spectrum (HD 98058) plus a cooler blackbody spectrum to simulate the thermal emission by the warm dust in the inner disk. The observed spectrum of HD 169142, the photosphere plus dust model, and the difference are shown, with the difference offset for clarity. \label{line-extractions}}
\end{figure}

\subsection{Characterization of the Spectra}

As noted by Grady et al. (2007), the FUV spectrum of HD 169142 is dominated by the stellar photosphere
for $\lambda\geq$ 1600 \AA, and then transitions to a continuum + emission-line spectrum at shorter
wavelengths. After smoothing by running boxcar filters, net  continuum fluxes in excess of 10$^{-15}$ erg/cm$^{2}$/s/\AA\ are present over the full spectral range of the COS data.  Following Grady et al. (2007),
we compare the smoothed COS data with a GHRS spectrum of Altair (A7Vn, V=0.76, B-V=0.22) scaled by $\Delta$V only (Figure 5).  The continuum in HD 169142 traces that of the scaled Altair data down to 1350 ~\AA , consistent with a photospheric origin. At shorter wavelengths, excess continuum light and emission line features are present.  The agreement between the scaled Altair spectrum and the COS data demonstrates that
there is no measurable selective extinction toward HD 169142 and is fully consistent with SED fitting discussed in $\S$3.3 and $\S$3.4. 

\begin{figure}[htpb]
\figurenum{5}
\epsscale{.6}
\plotone{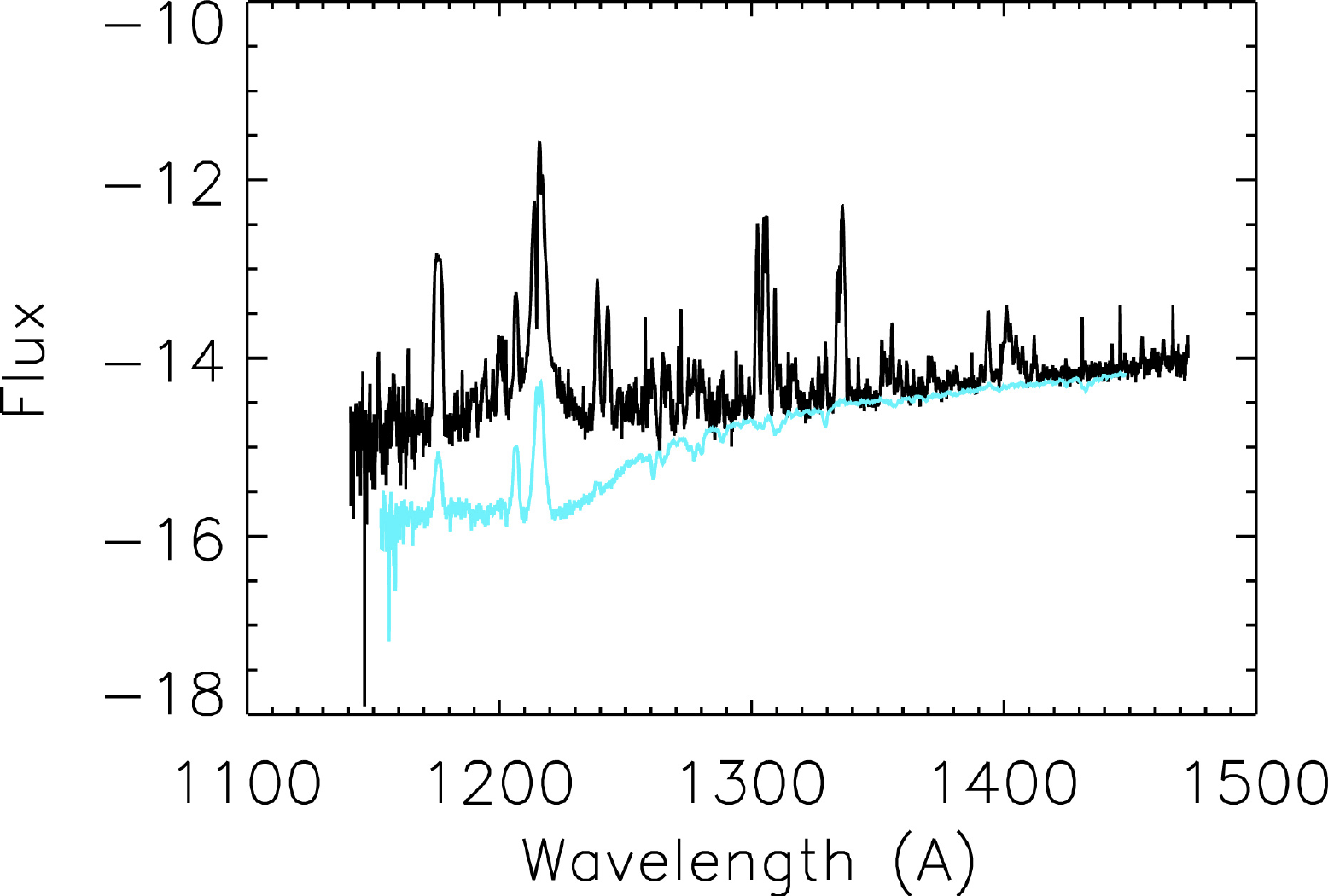} 
\caption{\footnotesize The smoothed COS data of HD 169142 (black) with a GHRS spectrum of Altair (A7Vn, V=0.76, B-V=0.22) scaled by $\Delta$V only (blue, lower spectrum).}
\end{figure}

\subsubsection{Lyman $\alpha$}

The strongest emission feature in the FUV spectrum of HD 169142 at all 3 epochs is Lyman $\alpha$,
similar to  other late A Herbig stars and classical T Tauri stars (France et al. 2014). 
Lyman $\alpha$ emission is detected at all 3 epochs of FUV spectroscopy (Figure 6), with varying amounts of 
geocoronal contamination affecting $\pm$200 km/s (COS data) with much less
contamination of the STIS data (within $\pm$40 km/s of line center for the G140M data). The observed emission profile consists of double-peaked emission, with wings extending to $\pm$900 km/s, and absorption which is not centered on the geocoronal emission or the stellar radial velocity. There is a factor of 2 change in flux from the April 2013 to July 2013 data, and $\approx 30\%$ variability from 2013 July 11 to 2013 July 13.  Variability at this level is similar to that previously reported for other well-studied Herbig stars such as HD 163296 and HD 104237. 

\begin{figure}[htpb]
\figurenum{6}
\epsscale{.6}
\plotone{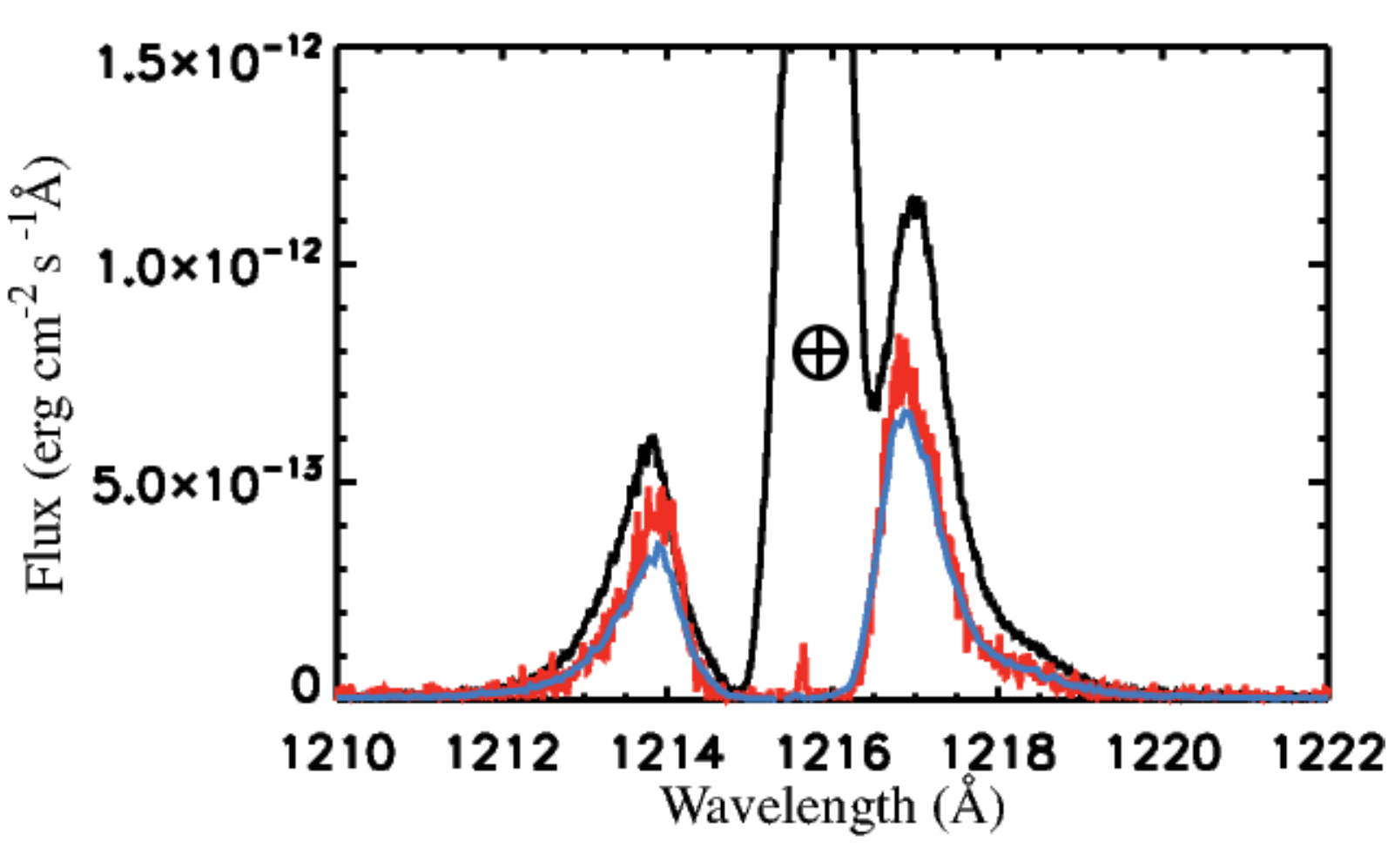} 
\caption{\footnotesize Lyman $\alpha$ emission at all 3 epochs of FUV spectroscopy, with varying amounts of 
geocoronal contamination centered on the rest wavelength. Data are colored-coded as 2013-04-05 COS (black), 2013-07-11 STIS E140M (red), and 2013-07-18 STIS G140M (blue). See the electronic edition of the Journal for a color version of this figure.}
\end{figure}

The STIS G140M data reveal a spatially unresolved source in both the continuum and in the emission
lines. The STIS data show saturated absorption in the geocoronal feature to +140 km/s, consistent
with interstellar and/or circumstellar H I absorption corresponding to N(H I)~2.5-5$\times 10^{19}$ cm$^{-2}$,
or E(B-V)=0.01 for gas to dust ratios typical of the diffuse ISM. None of the Lyman $\alpha$
profiles show a sharp emission feature similar to the jet emission seen in HD 163296 or HD 104237,
so we interpret the blue shifted absorption as evidence for a stellar wind, but not a jet. We measure
the terminal velocity of the wind at the 50\% of peak blue emission level as ranging from -364 to -525 km/s. 

\subsection{Modeling the Post-2000 State of the Disk}

In investigating the structures within the protoplanetary disk in HD 169142, we present several distinct models corresponding to the two states of the SED. Since high contrast and angular resolution imagery dating from the pre-2000 era is unavailable, we explore both states in reverse chronological order, beginning with modeling the post-2000 state of the broadband SED and multi-band imagery, which we describe in detail in this subsection. Later, we adapt this model into a set of three models which match the pre-2000 SED and explore several speculative scenarios for disk structures at sub-AU radii. 

\begin{figure}[h]
\epsscale{.8}
\figurenum{7}
\plotone{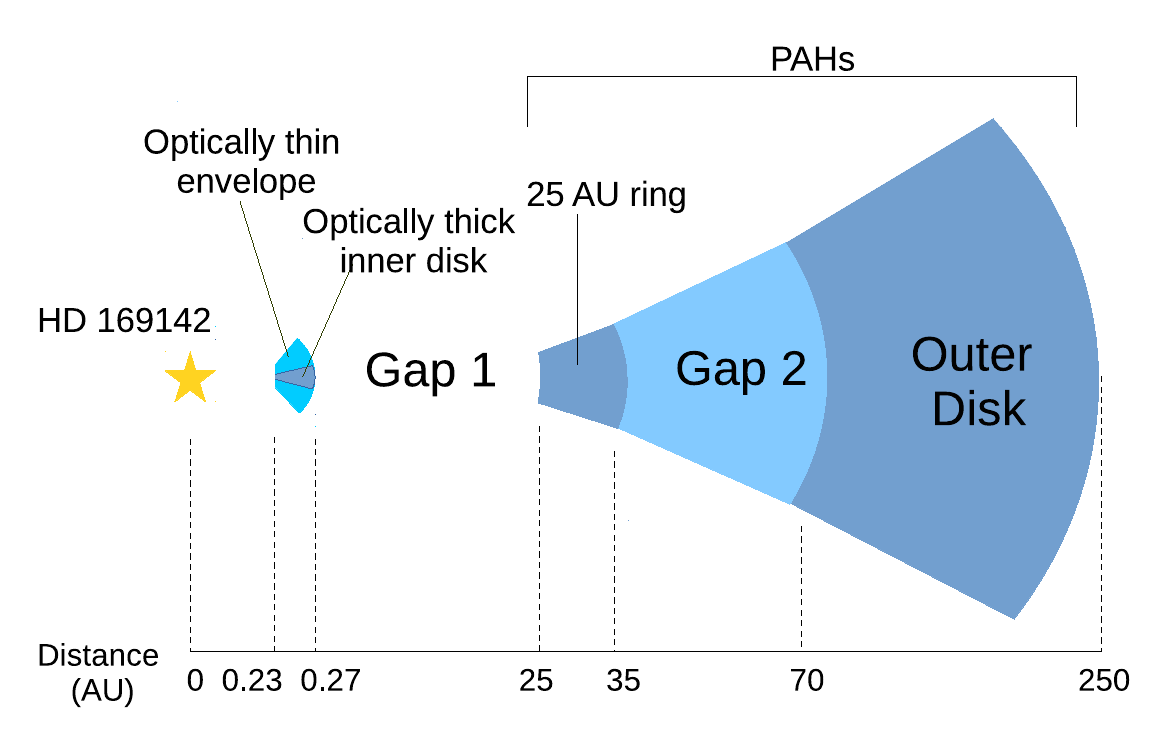} 
\caption{\footnotesize Drawing detailing the various model disk components as used in our model of the post-2000 state (only the sub-AU regions are different in our pre-2000 models).\label{diskcomp}}
\end{figure}

The Hochunk3D software allows for two separate disks (each with two coexisting grain populations), one gap (with its own density and scale height profiles), and an envelope. Usually, one disk is dedicated to small grains and the other dedicated to large, settled grains -- usually with coplanar geometries. However, to model a dual-gapped nature of HD 169142, the available disks were constructed as follows: The first disk was adopted for the optically thick inner disk beginning at 0.23 AU and ending at 0.27 AU. The second disk was adopted as the 25 AU ring and outer disk, with the second gap placed between 35-70 AU.\footnotemark \footnotetext{With this construction and the available disk structures in the software, we were not able to include in our models the large grain mid-plane disk outward of 25 AU. While this construction is adequate for our purposes of modeling the SED and NIR imagery, the addition of more disk structures to currently available software will be important for future studies to explore the distribution of material within the disk in greater detail.} \footnotemark \footnotetext{We have used 35 AU as the inner radius of the second gap in our models as this provides a better match to the integrated flux percentages and comparable size of the ring and the gap in the images than our models with the inner radius of the gap at 40 AU.} Figure 7 details the location of the disk components in the post-2000 model, whose SED is presented in Figure 8. The full scope of the parameter space explored and the values used in construction of the post-2000 model are detailed in Appendix A. Where available, parameter values were adopted from previous studies, indicated in table A1 by their references.

\begin{figure}[htbp]
\figurenum{8}
\epsscale{.8}
\plotone{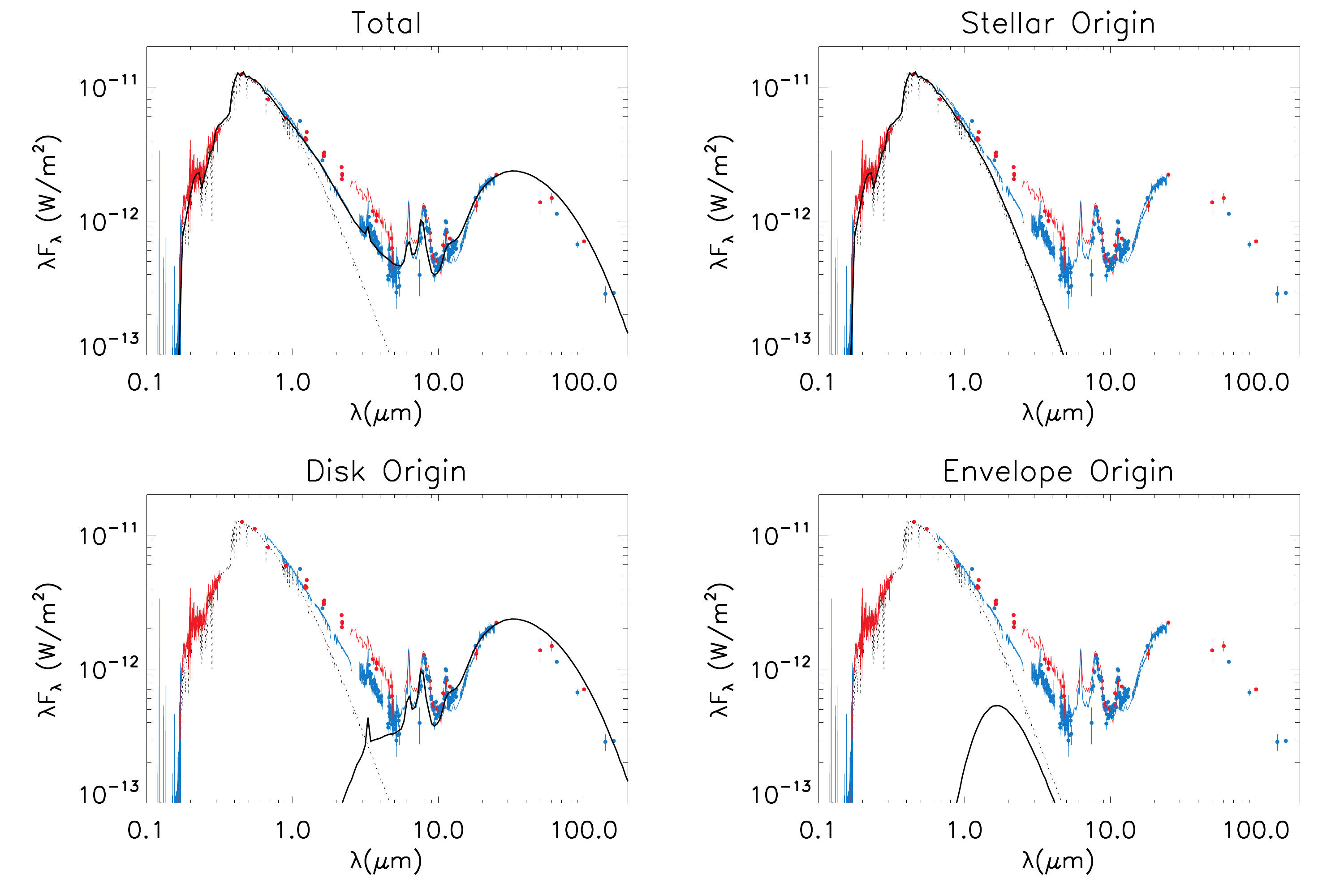} 
\caption{\footnotesize HD 169142 post-2000 model spectral energy distribution (solid black). Pre-2000 observations (high state) are presented in red, while post-2000 observations are presented in blue. The dotted black line represents the Kurucz 7500 K stellar atmosphere model. See the electronic edition
of the Journal for a color version of this figure.}
\end{figure}

In attempting to fit the SED and imaging available for HD 169142, we place constraints on several disk geometries and structures. First, models without an inner disk fail to reproduce the observed NIR flux levels. This requires the presence of material close to the central star, as suggested by \cite{osorio14} and others that the protoplanetary disk around HD 169142 belongs to the class of objects known as pre-transitional disks \citep{espaillat07}. For the outer disk grain population, we have used model 1 from \cite{wood02}, which includes a mixture of amorphous carbon and astronomical silicates with a maximum grain radius of 1 mm, but which does not produce a strong 10 $\mu$m silicate emission band. To remain consistent with an ``inside-out" profile of grain growth (see \cite{tanaka05}), we have used model 3 from \cite{wood02} for the mid-plane of our inner (sub-AU) disk, which similarly has a maximum grain radius of 1 mm, but with a size distribution resulting in a population of larger grains than in the outer regions of the disk. The emission from PAHs in the disk do not exhibit a variable emission profile similar to the underlying continuum between 1.5-10 $\mu$m (see Spitzer/IRS and ISO/PHT-40 data, Figure 3), suggesting little to no PAHs within the variable inner disk. Hence we include these molecules in the outer (25-250 AU region).\footnotemark \footnotetext{Given the available structures in the code, this is as close as we could come to matching reported locations of PAH emission in \cite{habart06} of 50\% of emission originating from within 29 AU and 90\% of emission originating from within 50 AU, suggesting that the largest concentration is within the $\sim$25-40 AU ring.} 

The envelope was adopted as a disk-like optically thin body of material suspended above and below the optically thick mid-plane of the inner disk (see Figure 7). In order to explore the source of the NIR variability, it was necessary that the changes between pre-2000 and post-2000 models occur primarily within this optically thin structure, as changes to the inner edge of the optically thick mid-plane produce variable shadowing of the outer disk, which is non-evident in the available data in the broadband SED.\footnotemark \footnotetext{Alternatively, changes to the optically thick mid-plane may not affect the shadowing of the outer disk in the case of a misaligned inner disk. However, such a scenario is ruled out as even relatively small inner disk inclinations produce minima in flux at a certain PA throughout the disk, which are not seen in the images.} Interestingly, an optically-thin composition of large grains (as expected of an ``inside-out" way of grain growth) fails to reproduce the amount of NIR emission without being placed within the dust sublimation radius. However, an optically-thin component of smaller grains (0.005-1 micron) succeeds at reproducing the observed emissions while remaining outside of the sublimation radius.\footnotemark \footnotetext{Specifically, we have used the same model from \cite{clayton11}, which uses the size distribution of Mathis, Rumpl, and Nordsieck (1977) and optical constants of burnt benzene from \cite{zubko96}.} The unexpected presence of these small grains at such altitude above the inner disk may hint at impacts or other debris-creating processes occurring within the inner regions of the disk, unless the grain growth and formation of the mid-plane is simply of very recent occurrence.

The profile of the outer disk emission very closely resembles the peak of a single temperature black-body of T=120 K, which is consistent with most of the emission originating from the inner wall of the outer disk, and allows us to place constraints on its structure. First, the SED is best fit by placing this wall at 25$\pm$2 AU, as suggested by previous modeling studies and imaging (\cite{honda12}, \cite{maaskant13}, \cite{quanz13}, \cite{osorio14}). Next, values are obtained for the A and B parameters, respectively the power law exponents for the radial density ($\propto r^{-A}$) and scale height ($\propto r^{B}$) of the outer disk, which largely impact the width and height of the emission profile. There exist several degenerate solutions for these parameters in combination with the disk mass. Out of these, the disk mass is most constrained by the SED. Disk masses $<$0.0015 M$_{\odot}$ fall short of reproducing the observed emissions at $\lambda \geq$60$\mu$m, while disk masses $>$0.003 M$_{\odot}$ produce too much emission at $\lambda \geq$60$\mu$m. A disk mass of 0.003 M$_{\odot}$ reproduces the 60 $\mu$m IRAS, 100 $\mu$m IRAS, and 160 $\mu$m AKARI/FIS, but not the 90 $\mu$m or 140 $\mu$m AKARI/FIS. Since the density of data in this portion of the spectrum is relatively low, and since the emissions at these wavelengths should not be significantly variable over the period of the observations, we have chosen to fit to the IRAS and 160 $\mu$m AKARI/FIS points, as M$_{disk}$ = 0.003  M$_{\odot}$ is consistent with the lower estimates of other studies (e.g. \cite{meeus10}, who obtained M$_{disk}=$0.005$\pm$0.002 M$_{\odot}$). However, we find that 0.0015 M$_{\odot} \leq$ M$_{disk} \leq$ 0.0025 M$_{\odot}$ may reproduce emissions consistent with the 90 $\mu$m and 140 $\mu$m AKARI/FIS points as well as the 50 $\mu$m photometry in \cite{harvey95}.

Since most of the flux between $\sim$15-40 $\mu$m is from the 25 AU wall, there exists another degeneracy between the scale height normalization and the flare of the disk, as only the height of the wall is most constrained by the SED. The H-band imagery and their corresponding surface brightness profiles allow us to explore beyond this degeneracy, as the regions outside of the 25 AU wall are resolved to have observable structure. Since the scale height normalization of the 25-250 AU structures is the same throughout (given our use of the available disk structures, see above), modifying the disk flare exponent in conjunction with the scale height normalization is our only means to adjust the height of the $\gtrsim$70 AU structures without affecting the height of the 25 AU wall. We find that the SED and images may be simultaneously reproduced with a disk flare $\propto$ $r^{1.3}$ for all structures where r$\geq$25 AU. Unfortunately, without additional disk structures being available in the 2014-01-31 release of the code, this eliminates our ability to explore the possibility that these regions may have independent geometries, or perhaps ones different than those necessary for our constructions (e.g. \cite{momose13} and \cite{grady07} suggest flares proportional to $r^{1.0}$ and $r^{1.065}$, respectively). Nevertheless, our models do maintain agreement with the SED, H-band surface brightness profile, and integrated disk flux percentages, and hence suggest that the values used in this study are entirely plausible.
	
\cite{momose13} reports a surface brightness profile behaving as $r^{-3}$ in the regions where r$<$52 AU and r$>$81 AU and suggest that a surface density profile proportional to $r^{-1}$ could reproduce the observed surface brightness.\footnotemark \footnotetext{The same suggestion is made by \cite{grady07} that a density profile proportional to $r^{-1}$ could reproduce the $r^{-3}$ surface brightness profile observed in HD 169142 by HST/NICMOS.} \cite{quanz13} reports a better fit to their data with a broken power law exponent of $-2.64^{+0.15}_{-0.17}$ for the region between 85-120 AU and an exponent of $-3.90^{+0.10}_{-0.11}$ for the region between 120-250 AU, suggesting the presence of additional discontinuities to the disk surface. Hence there is no exact agreement between the observations, although an approximate $r^{-3}$ fall off seems well established between 80-120 AU. While maintaining fits to the SED, we find that a disk flare $\propto$ $r^{1.3}$ and density profile at the mid-plane\footnotemark \footnotetext{Note that this density profile is different than the surface density profile which is integrated over the vertical scale height of the disk.} $\propto$ $r^{-1}$ reproduce an H-band surface brightness profile of $r^{-2.6}$ between 80-200 AU, while we may reproduce a steeper surface brightness profile $\propto r^{-3}$ with a steeper fall off to the mid-plane density $\propto r^{-2}$. Given the available disk structures in the code, we were not able to explore additional discontinuities in the disk surface to match a steeper fall off outward of 120 AU. Hence we have chosen to match the surface brightness profile of the brightest inner portion by including the $r^{-1}$ radial density profile between 25-250 AU in the final form of our post-2000 model. Note that as a result our model image which appears later in the article (Figure 10, center) appears brighter in the outermost regions of the disk than the \cite{quanz13} image with the same scaling of intensity by $r^{2}$. Since the surface brightness profile between $\sim$40-70 AU requires a separate in-depth treatment of this region, we present our complete model surface brightness profile later in Figure 11.
	
\subsection{Modeling The Second Gap}

One of the most striking features present in the multi-band images of the disk in HD 169142 is a second dark ring located between $\sim$40-70 AU (\cite{momose13}, \cite{quanz13}, \cite{osorio14}, in addition to the $\sim$0.3-25 AU gap that is heavily discussed in the literature (e.g. \cite{grady07}, \cite{honda12}, \cite{maaskant13}). Such a feature in the NIR images may be either the result of shadows cast by the wall at $\sim$25 AU or the signature of a second gap -- one which is not so devoid of material as the gap inward of 25 AU as $\sim$14\% of the integrated disk flux emanates from this region \citep{quanz13}. Using the VLA, \cite{osorio14} detect 7 mm emission concentrated between $\sim$25-40 AU, although much more asymmetric than the NIR emission ring. As the 7 mm emissions penetrate deep within the disk detecting the thermal emissions of large dust grains, this result indicates that this is a true density feature which extends through the mid-plane of the disk. Hence any shadows cast by inner structures must be in addition to effects caused by the density feature. 

To explore the extent that shadows may be contributing to the NIR dark ring, we present three separate models. The first model does not contain any region-specific modifications to the density or scale height profile in the region of the second gap. This model is in attempt to determine the scale height of the 25 AU wall and other structures which are constrained by the SED, and to see whether or not the second dark ring appears naturally as a shadow in the model images. Although this model reproduces the observed SED, no second dark ring is present in the images (Figure 9, left). Thus, we explore the density conditions which may reproduce this feature. We find that the integrated H-band flux percentages (49\% bright ring - 14\% dark ring - 37\% outer disk) reported in \cite{quanz13} may be reproduced to within 2\% by scaling the density between 35-70 AU by 0.3$\times$ the value just outside of 70 AU (Figure 9, center). However, we also find that decreasing the scale height of the 35-70 AU region by 0.86$\times$ may also reproduce the observed NIR integrated flux percentages and radial brightness profile. Hence there may exist a multitude of models fitting the NIR images with a gap density change of $\geq 0.3\times$ and a scale height change of $\geq 0.86\times$. However, lacking the correct density structure at the mid-plane, many of these models would fail to reproduce the observed lack of 7 mm emission from the gap (see Figure 9, right). This degeneracy too may be overcome by modeling to fit the NIR and 7 mm images simultaneously, although this is beyond the scope of our study at this stage. Similarly, future acquisition of multi-band imagery will help to constrain the density and location of various grain species throughout the disk. We leave the density/scale height degeneracy to be resolved in future studies and include only a density scaling to the gap in our final model. 

\begin{figure}[htbp]
\figurenum{9}
\epsscale{.6}
\plotone{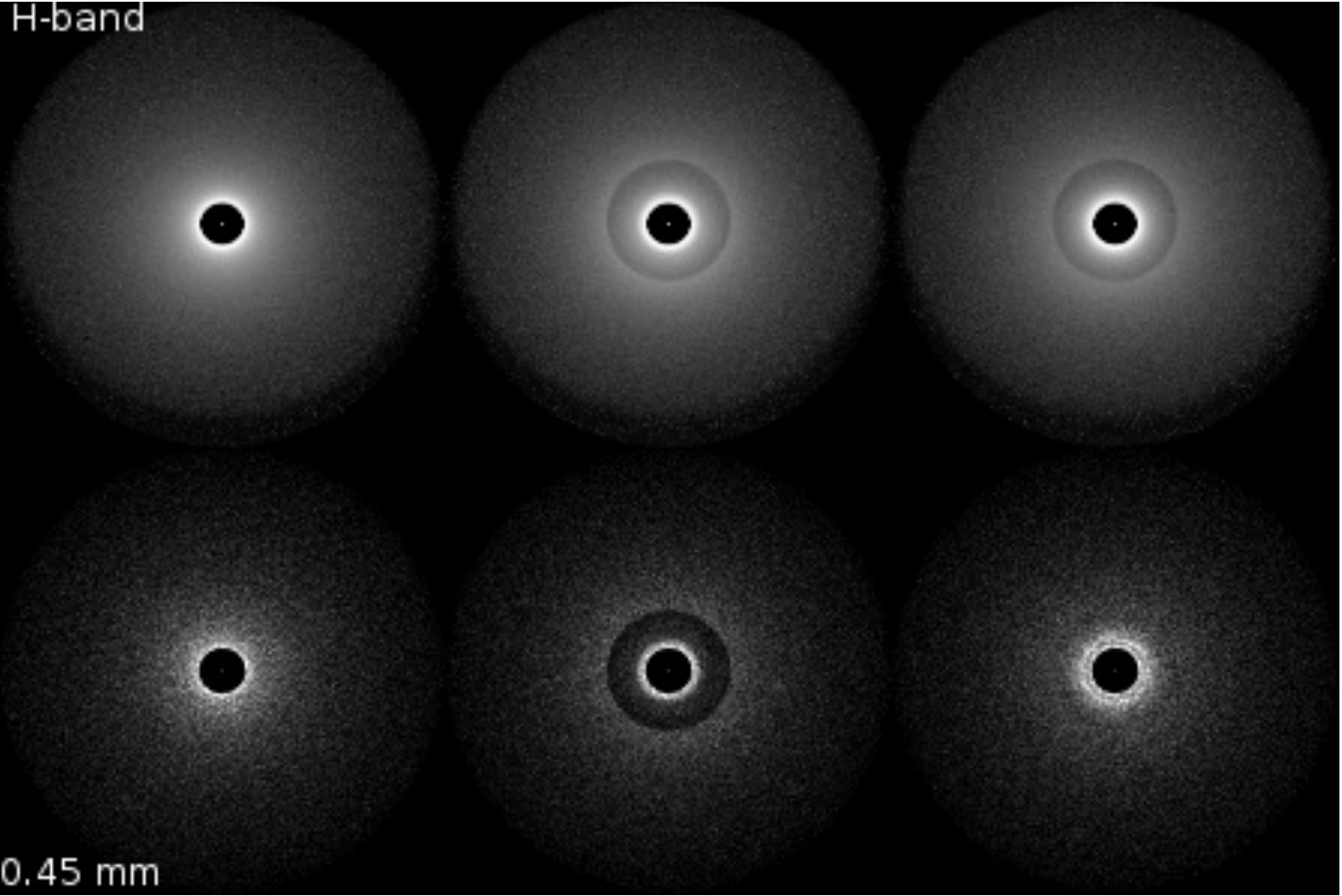} 
\caption{\footnotesize Presented are model H-band (top, scaled as $log I$) and 0.45 mm images (bottom). $\textbf{Left:}$ post-2000 model with no density / scale height discontinuity between 35-70 AU.   $\textbf{Center:}$ post-2000 model with 0.3$\times$ density scaling between 35-70 AU.  $\textbf{Right:}$ post-2000 model with 0.86$\times$ adjusted scale height between 35-70 AU. Note that the small dot in the center of all of the images is the emission from the inner (sub-AU) disk. \label{gaps}}
\end{figure}


\begin{figure}[h]
\epsscale{.7}
\figurenum{10}
\plotone{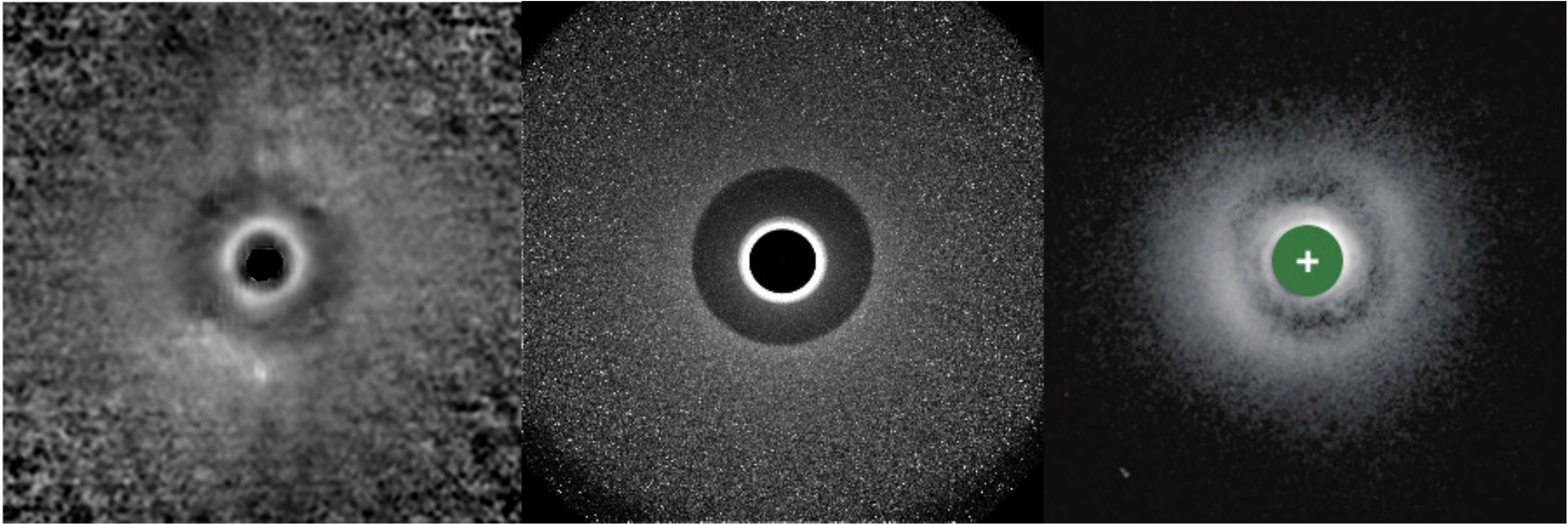} 
\caption{\footnotesize $\textbf{Left:}$ VLT/NACO H-band polarized flux image (Figure 1, \cite{quanz13}) intensity scaled by $r^{2}$. $\textbf{Center:}$ H-band polarized flux post-2000 model image (intensity scaled by $r^{2}$). $\textbf{Right:}$ Subaru/HiCIAO H-band polarized flux image (Figure  1, \cite{momose13}). Each image has the same scale of 400 AU across. \label{images}}
\end{figure}

Furthermore, we find that a density profile $\propto r^{-1}$ in the gap produces a surface brightness profile in the region following a power law of $\sim r^{-2}$, as opposed to the observed profile of $r^{-3}$ reported in \cite{momose13}. Since a significant amount of radiation is back-scattered off of the 70 AU wall, we find that a steeper density profile $\propto r^{-2}$ inside of the gap is necessary to reproduce the $r^{-3}$ surface brightness fall off. Thus our final H-band (2.2 $\mu$m) model image of the post-2000 state (Figure 10, center) highly resembles those taken through the actual telescopes with a surface brightness profile (Figure 11) which is closely described by a power law of $r^{-3}$ where r$<$50 AU and $r^{-2.6}$ where r$>$80 AU. 

\begin{figure}[h]
\epsscale{.75}
\figurenum{11}
\plotone{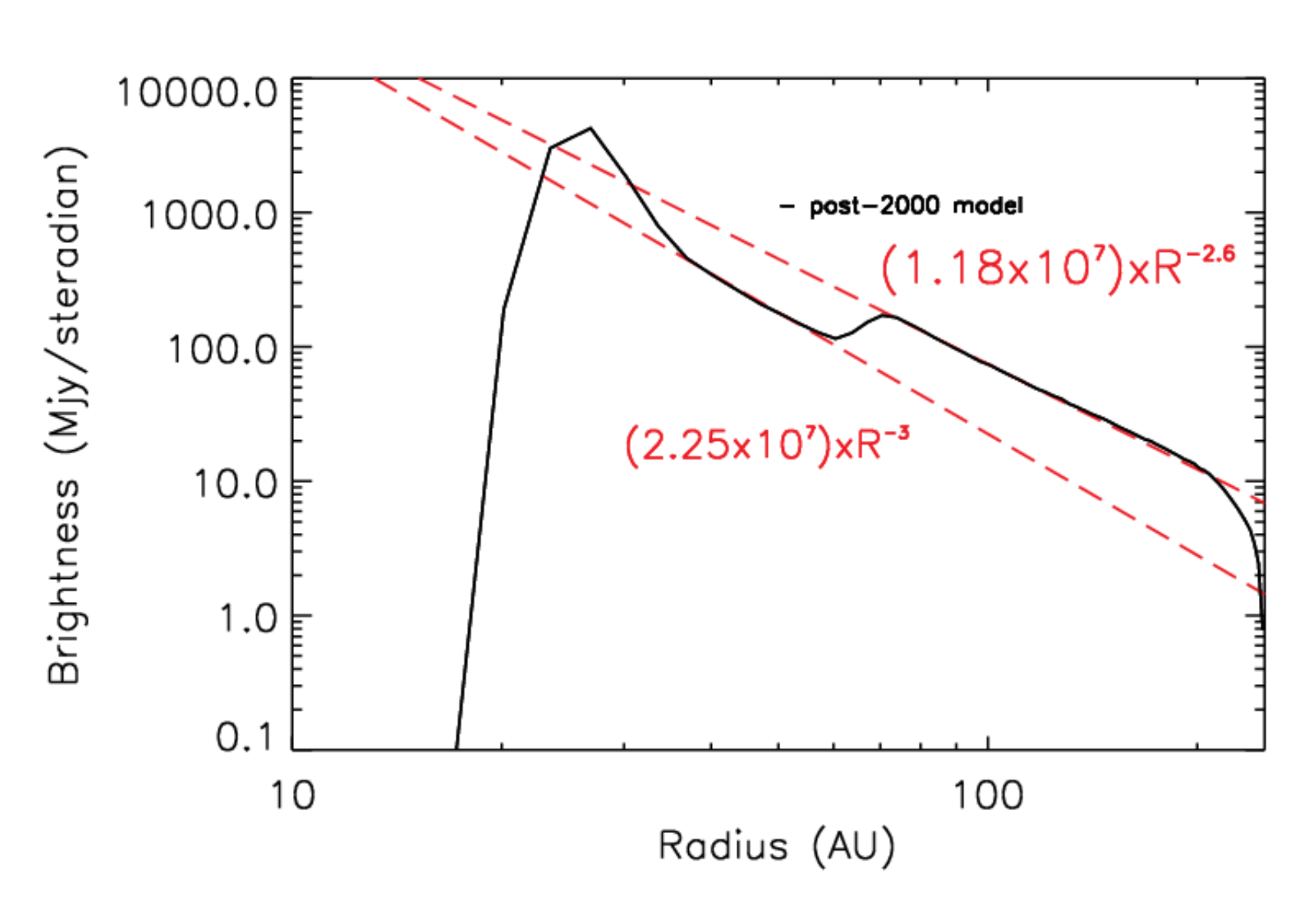} 
\caption{\footnotesize Azimuthally averaged H-band polarized surface brightness of our post-2000 model, with power law models in red (dashed). \label{radialbrightness}}
\end{figure}		

\subsection{NIR Variability and Structures at Sub-AU Radii}

As detailed in section 2.5, our construction of the SED reveals variability in the NIR spectrum, specifically at wavelengths between 1.5-10 $\mu$m. Measurements fall into two distinct flux states, with those recorded after the year 2000 exhibiting up to 45\% lower NIR flux than those recorded prior to 2000, with the change occurring over no more than 10 years (see Figure 3).

We present three models corresponding to the pre-2000 state of the disk in HD 169142. Model A explains the NIR variability by changing the location of the outer edge of the inner disk, model B by changing the altitude and density of the mini-envelope, and model C by the variable density of the mini-envelope alone. 
We have restricted changes in model structures to the inner regions due to the difficulty for changes to propagate through the outer structures faster than their orbital periods, which would certainly be longer than the observed timescale of $\sim$ 10 yr. Note that changing the location of the outer edge of the optically thick mid-plane does not affect the shadowing of the outer disk, and contributes to the NIR variability since it is being heated by the optically thin envelope. It is also interesting to note that even in our most extreme scenarios, the change to the envelope mass accounts for less than 1\% of the mass lost due to accretion ($\approx 2\times 10^{-8} M_{\odot}$ over the maximum timescale of 10 years). Hence the optically thick mid-plane is a necessary component of our model constructions to explain the accretion rate. For comparison, the cross-sectional density distribution of the sub-AU regions of each model are presented in Figure 12, which we now discuss. 

\begin{figure}[htpb]
\epsscale{1}
\figurenum{12}
\plotone{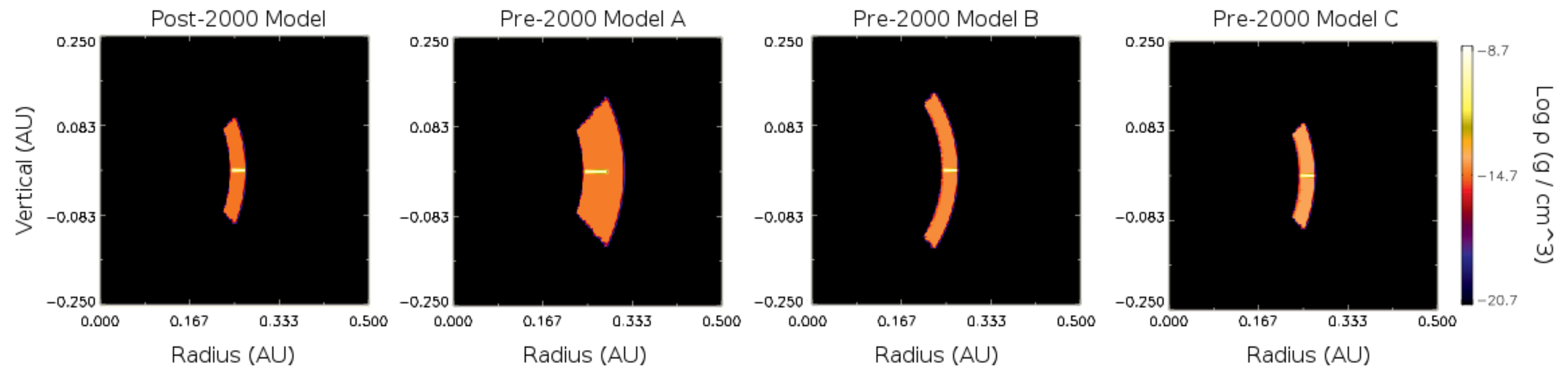}
\caption{\footnotesize Cross-sectional density distribution of sub-AU model structures. Note the small color difference between post-2000 and pre-2000 model C envelopes corresponding to the 4$\times$ density difference, which may be difficult to see with the logarithmic scale. See the electronic edition of the Journal for a color version of this figure.}
\end{figure}

\subsubsection{Pre-2000 Model A}

In this model the pre-2000 NIR SED is achieved through extending the outer edge of the inner disk and mini-envelope to 0.285 AU and 0.32 AU, respectively (see model SED in Figure 13 and density profile in Figure 12). Chronologically from this model to the post-2000 model, these correspond to an inward shift of 0.015 AU for the optically thick disk and 0.05 AU for the optically thin envelope. Changes analogous to these might occur in systems in which the inner disk is actively being cleared of material by accretion onto the central star and/or through gravitational sculpting by a planetary mass companion orbiting within the gap (see \S 4.2). To keep the density of the inner disk consistent with that of the post-2000 model, the fraction of mass in the inner disk was adjusted accordingly to 7.1$\times 10^{-5}$ M$_{disk}$, as opposed to the value of 5.0$\times 10^{-5}$ M$_{disk}$ in the post-2000 model. These correspond to a net change in mass for the mid-plane of 6.3$\times 10^{-8}$ M$_{\odot}$ and for the envelope of 4.0$\times 10^{-11}$ M$_{\odot}$. 

\begin{figure}[h]
\figurenum{13}
\epsscale{.9}
\plotone{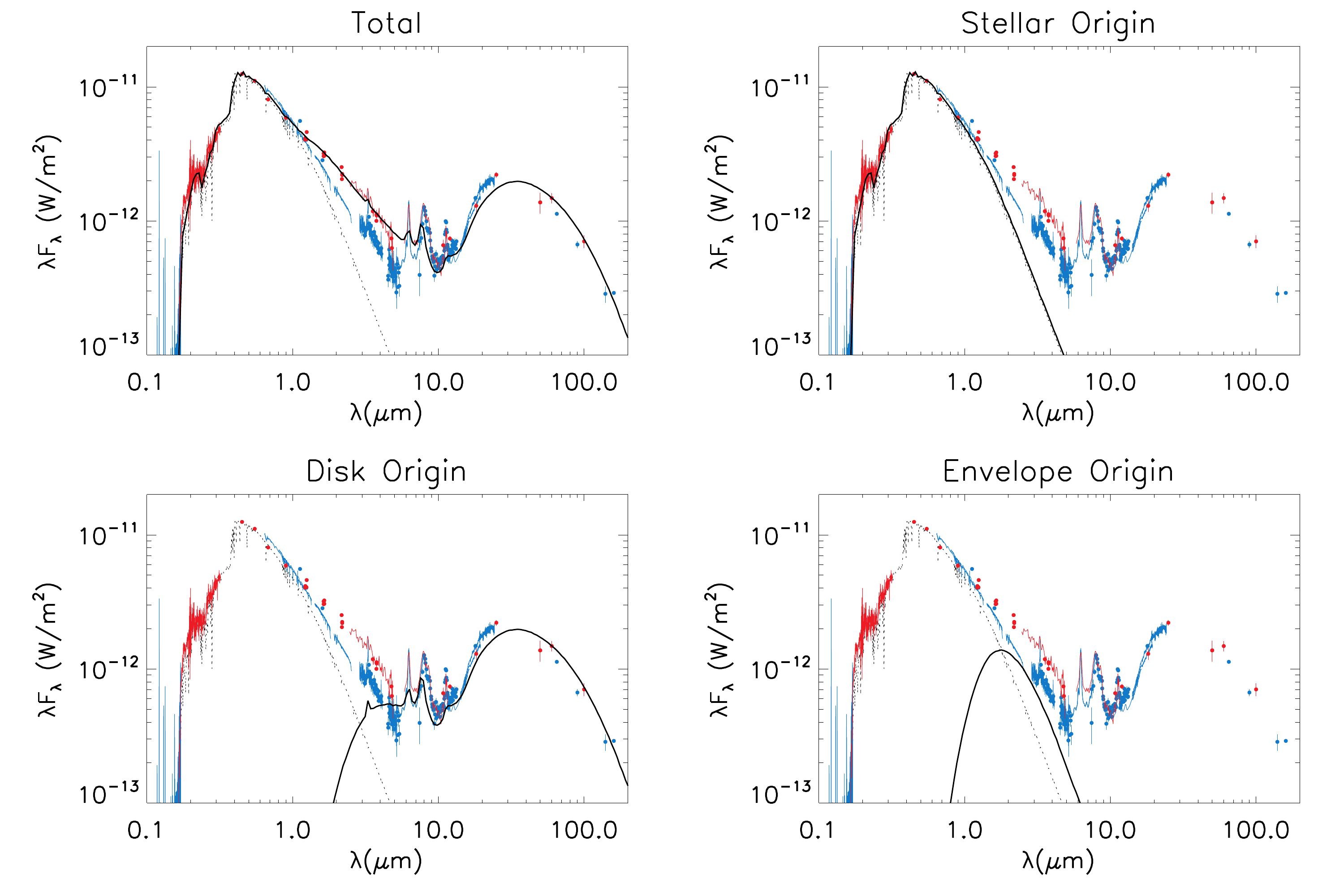} 
\caption{\footnotesize HD 169142 pre-2000 model A spectral energy distribution (solid black). Pre-2000 observations (high state) are presented in red, while post-2000 observations are presented in blue. The dotted black line represents the Kurucz 7500 K stellar atmosphere model. See the electronic edition of the Journal for a color version of this and subsequent figures.}
\end{figure}

\subsubsection{Pre-2000 Model B}

In this model the pre-2000 NIR state is achieved through increasing the scale height of the mini-envelope by +0.07 AU and doubling its density to a uniform 5$\times$10$^{-15}$ g cm$^{-3}$, as compared to the post-2000 model (see model SED in Figure 14 and density profile in Figure 12). These changes correspond to a mass loss of 6.6$\times 10^{-11}$ M$_{\odot}$ from the envelope in going from this model to the post-2000 state. To account for the mass lost due to accretion of $\sim$2$\times 10^{-8}$ M$_{\odot}$ (obtained using our derived accretion rate over 10 years), we have adjusted the fraction of mass in the inner disk to 5.58$\times 10^{-5}$, as opposed to the value of 5$\times 10^{-5}$ used in the post-2000 model. The magnitude of an altitude change alone necessary to reproduce the observed flux levels would effectively constitute a halo in the high flux state, and its unlikely near-complete disappearance in the low flux state. Hence in this model we have chosen to vary both the altitude and the density of the mini-envelope simultaneously. Changes analogous to these might occur in systems in which material is suspended at higher altitude in the pre-2000 state as opposed to the post-2000 state, perhaps due to an inner disk wind or other phenomena within the inner regions (see \S 4.2). 

\begin{figure}[tbp]
\figurenum{14}
\epsscale{.9}

\plotone{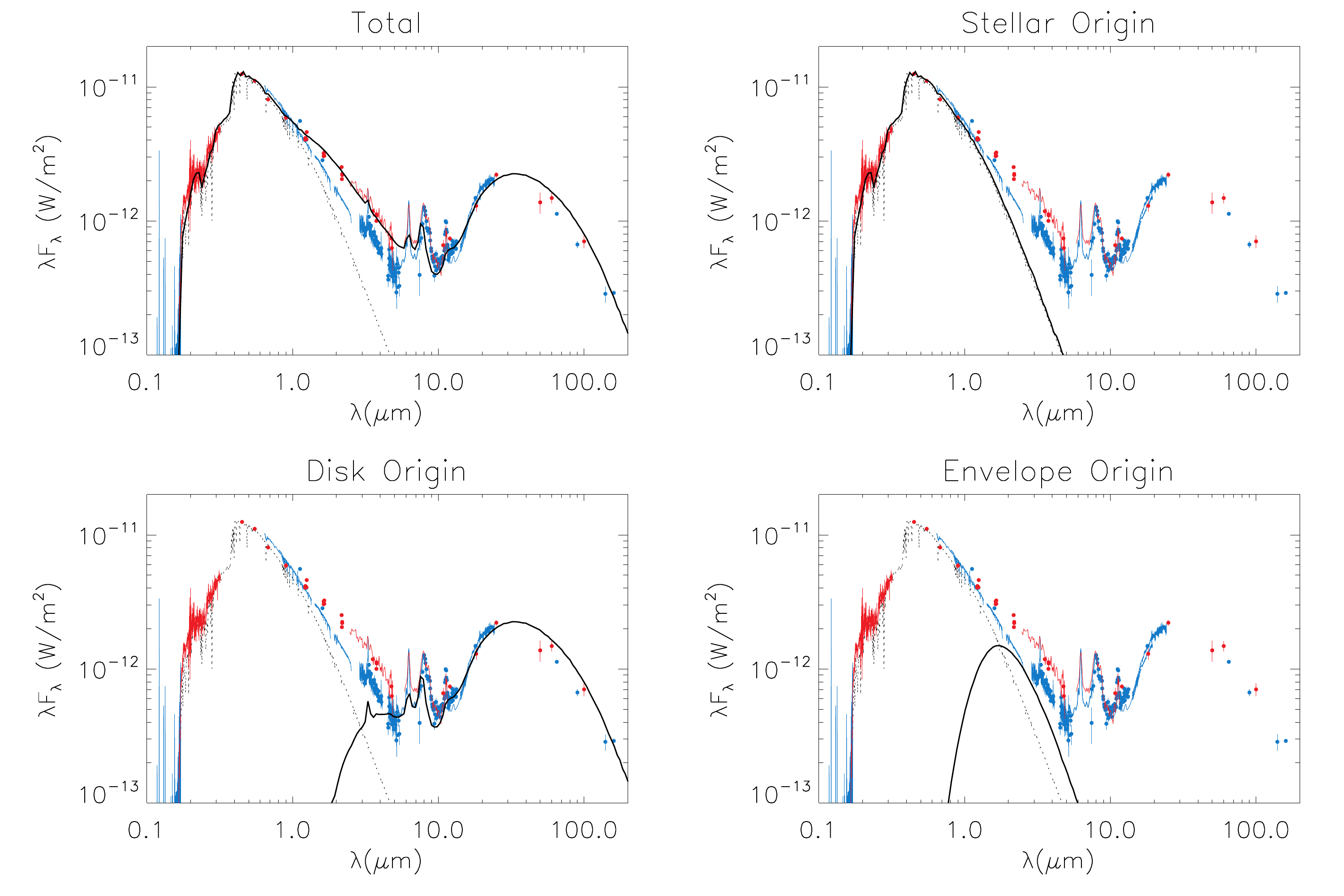} 
\caption{\footnotesize HD 169142 pre-2000 model B spectral energy distribution (solid black). Pre-2000 observations (high state) are presented in red, while post-2000 observations are presented in blue.}
\end{figure}

\subsubsection{Pre-2000 Model C}

In this model the pre-2000 NIR state is achieved through increasing the density of the mini-envelope alone to a uniform 10$^{-14}$ g cm$^{-3}$, which corresponds to a mass loss of 7.2$\times 10^{-11}$ M$_{\odot}$ (see model SED in Figure 15 and density profile in Figure 12). Again, to account for the mass lost due to accretion of $\sim$2$\times 10^{-8}$ M$_{\odot}$, we have adjusted the fraction of mass in the inner disk to 5.67$\times 10^{-5}$, as opposed to the value of 5$\times 10^{-5}$ used in the post-2000 model.  While this change in density may plausibly be due to accretion onto the central star, such a dramatic change in density to the structures at high altitude produces variable shadowing (and thus emission) of the outer disk. We present this model solely as a means of narrowing down where and how the changes must be taking place within the sub-AU structures.

\begin{figure}[htpb]
\figurenum{15}
\epsscale{.9}
\plotone{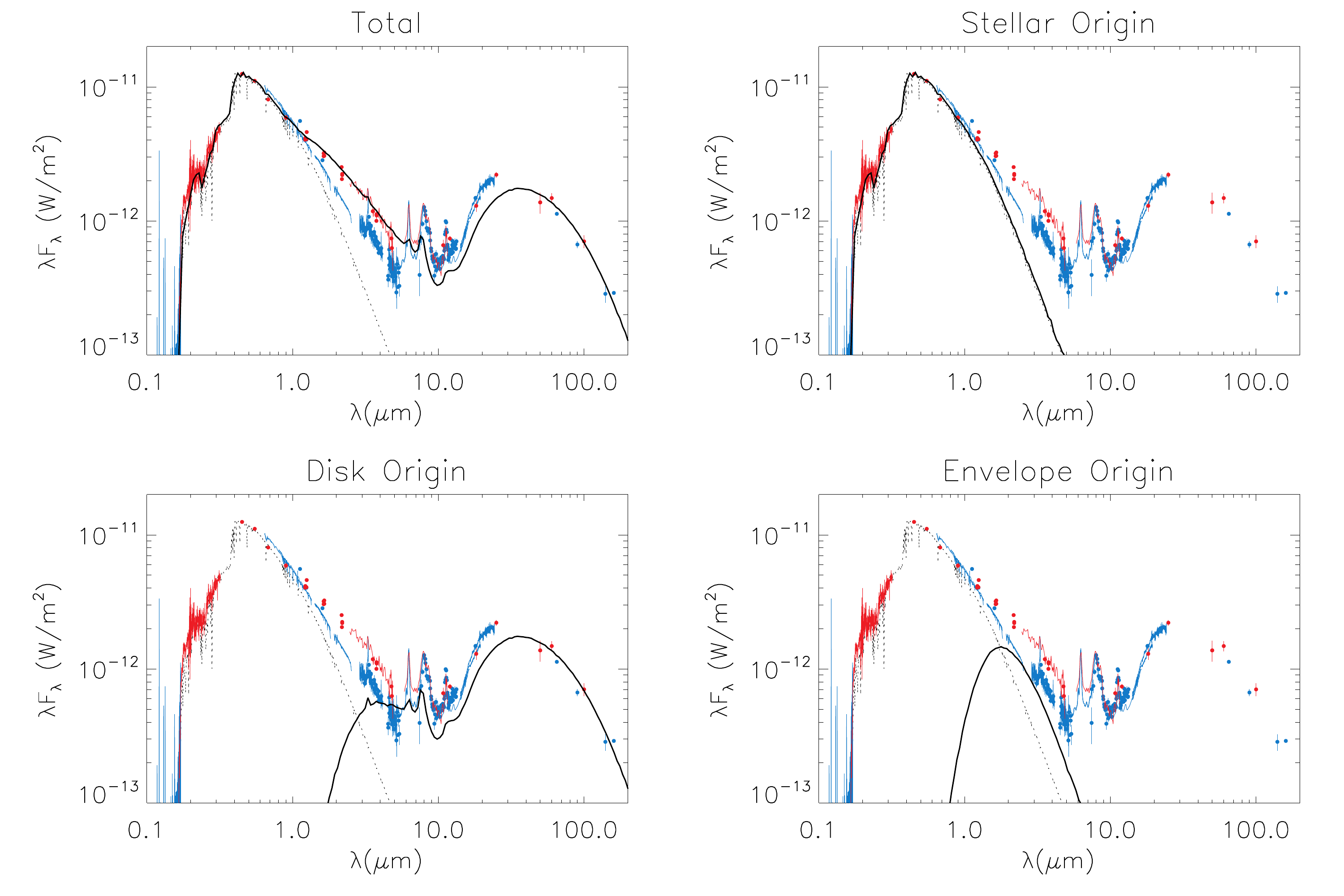} 
\caption{\footnotesize HD 169142 pre-2000 model C spectral energy distribution (solid black). Pre-2000  observations (high state) are presented in red, while post-2000 observations are presented in blue.}
\end{figure}

\section{Discussion}

\subsection{Summary of Results}

Our primary finding is that the protoplanetary disk in HD 169142 possesses material that is within 1 AU the star and undergoing changes to its structure on the order of a decade. A simple way to reduce the NIR flux from observed pre-2000 levels to post-2000 levels is to reduce the scale height of the innermost edge of the disk. However, such a change would decrease the shadowing of the outer disk, and hence produce an anti-correlated variability of the inner and outer disk emissions, which is non-evident in the SED. Hence, we have included in each of our models an optically thin population of small grains enveloping the optically thick sub-AU mid-plane. Changes to this optically thin structure do not produce major variations in the shadowing of the outer disk. The presence of such a structure is consistent with the results of \cite{dullemond10} who find that more realistic models of inner dust rims are insufficient to match ``bump"-like NIR emission, as well as the results of \cite{vinkovic14} who suggests that the standard puffed-up inner disk rims are insufficient to explain the NIR excess of many Herbig Ae and related stars and that additional sources of NIR emission must be present. 

We find three scenarios for the sub-AU structures in the pre-2000 (high NIR flux) state, one of which we may almost entirely exclude (model C) as it produces variable shadowing of the outer disk. We find nothing in the data to prefer either of the two remaining plausible scenarios over the other, although the changes between pre-2000 B $\rightarrow$ post-2000 are much more dramatic than the changes between pre-2000 A $\rightarrow$ post-2000. In the scenario of our pre-2000 model A $\rightarrow$ post-2000 model, the NIR variability is explained through an inward shift of the outer edge of the sub-AU structures toward the central star -- through accretion onto the star and perhaps other means. In the other scenario (model B $\rightarrow$ post-2000 state), the NIR variability is explained through the continual vertical evacuation of material from the sub-AU structures, leaving the post-2000 sub-AU structures with a smaller vertical scale height than those structures in the pre-2000 state. While these two scenarios accurately reproduce observed pre-2000 NIR flux levels through physically plausible mechanisms, we stress that the SED is highly degenerate to other structures and there exists a noteworthy potential for other presently unimagined possibilities.

Furthermore, we find that the dark ring observed between $\sim$40-70 AU in the polarized flux images is not the result of shadows cast by inner disk structures which are well constrained by the SED. Instead, we find that this feature may be reproduced in the model images through scaling the density of the region by 0.3$\times$ the value just outside of the gap, which is consistent with the density structure inferred through the 7 mm observations of \cite{osorio14}. However, we find that the H-band images may also be reproduced by adjusting the scale height of the region by 0.86$\times$ the value just outside of the gap. Hence, further study of the multi-band imagery is necessary here to eliminate the degeneracy of these parameters and determine the true composition of the various dust grain sizes throughout the disk.
\subsection{Origins of Structures}

\subsubsection{The Inner Disk and the R$\leq$25 AU Gap}

Extending to a distance comparable to between the orbits of Uranus and Neptune in our own Solar system, the $\sim$0.3-25 AU gap in the protoplanetary disk in HD 169142 is temptingly explained through the clearing of material by planetary mass companions. However, there are other mechanisms which are likely to produce the inferred disk structure and which must be accounted for before we may speculate on which (if any) features may be the signature of young and forming planets. Additionally, the timescale on which these mechanisms act to clear regions of the disk is of importance as the potential for planetary formation is significantly diminished with the decreasing supply of material. In a general overview of the evolution of protoplanetary disks, \cite{williams11} suggest that viscous accretion first dominates over photo-evaporation, until the mass accretion rate drops to $\sim 10^{-10}-10^{-9} M_{\odot} yr^{-1}$. Once these two counteracting processes become active on the same order of magnitude, the viscous flow of material from the outer disk will no longer be able to resupply the inner disk of material lost due to photo-evaporation, and an inner hole will be cleared. Through analysis of the Pa $\beta$ and Br $\gamma$ lines present in the May 2013 SpeX observations we derive a mass accretion rate of \.{M} $\approx$ $(1.5-2.7)\times 10^{-9} M_{\odot} yr^{-1}$. Hence this accretion rate suggests that such a scenario may be an active clearing mechanism for the inner ($<25$ AU) gap. Should this be the case, we may consider these to be ongoing processes within the presently observed state of the disk, as NIR flux levels (as of 2013) demand for the presence of material at sub-AU radii. Such a scenario is consistent with our speculative changes responsible for producing the variable NIR emission (see pre-2000 model A, in which the decrease in NIR flux is explained through the shift of material inward toward the star, with a net mass loss of 6.3$\times 10^{-8}$M$_{\odot}$). We may also assume that the disk clearing mechanisms of grain-growth and coagulation into larger bodies, as well as large-scale sculpting by newly formed planets, may also be responsible for producing these observed structures. In fact, the results of our pre-2000 model A are suggestive that such additional clearing may plausibly be occurring, as our derived mass accretion rate accounts for as much as 2.7$\times 10^{-8}$M$_{\odot}$, or 43\% of the total mass cleared over the observed timescale of the decrease in flux. This would leave room for other mechanisms to be actively clearing the region on the same order of magnitude as accretion onto the star.\footnotemark \footnotetext{Although it should be noted that in this model we have made the arbitrary assumption that the density of the midplane is the same as in the post-2000 state. If the density of the mid-plane were allowed to increase between pre-2000 and post-2000 states then the amount of mass lost between the two models may be brought down to the estimated amount of mass lost due to accretion.} Should a scenario such as this actually be underway, we could perhaps expect to see complete disappearance of the inner disk in HD 169142 on the order of another decade or so, prompting a change of class from pre-transitional to transitional disk.


While the NIR variability may be explained through the above scenario, there exists potential for other possibilities. Another scenario in which the NIR variability may be reproduced is one in which the optically thin envelope extends to higher altitude in the pre-2000 state than in the post-2000 state, as explored in our pre-2000 model B. This scenario matches pre-2000 NIR flux levels with a shifting of the envelope by 0.07 AU in the vertical direction and doubling its density as compared to the post-2000 model. While determining the exact mechanism and situations for which such a scenario may be relevant are beyond the scope of this paper, one likely possibility is that reconnection events (like those present in the model of \cite{shu94}) may be occurring within the magnetic field of a partially ionized inner disk, resulting in material which is lofted at altitude above the mid-plane. Such proceedings would also help to explain the presence of the optically thin material which has been necessary throughout our constructions. Similar considerations were explored for HD 163296 and HD 31648 in \cite{sitko08}, and the migration of material at altitude above the inner disk was explored as a mechanism for producing the photometric variability observed in HD 163296 \citep{ellerbroek14}. However, given the chronology of the observed states, such a scenario would have material decreasing in altitude by 0.07 AU and decreasing in density by a factor of 2 between the pre-2000 state and the post-2000 state. These changes, as well as those between our pre-2000 model C and post-2000 model are much more dramatic than those between the pre-2000 A and post-2000 models, and thus by invoking Occam's razor we consider the chronology of our pre-2000 model A $\rightarrow$ post-2000 model as being the most favorable of our presented scenarios. 
	
\subsubsection{The Second Gap}

We obtain a null result for the dark ring between $\sim$40-70 AU being a shadow cast by inner structures, but rather we find that this feature may be reproduced in the model images by a net deficit in the density of scattering material in the region. Hence we are left with the problem of how such structure in the disk might have originated and evolved from the primordial disk. Again, large-scale sculpting by planetary mass companions is temptingly invoked as a mechanism for clearing this gap of material -- especially given its location beyond the snow-line and the large amount of gas available here for giant planet production. Another possibility is that grain growth and coagulation may have been active within the region -- resulting in a net decrease in the density of micron sized grains and a corresponding increase in the density of grains with millimeter cross-sections. However, this seems unlikely as the 7 mm emissions suggest that the largest grains are concentrated within the $\sim$25-40 AU region, and not within the gap \citep{osorio14}. This result is also consistent with a planet-induced dust filtration process (as in \cite{zhu12}) being responsible for clearing of this gap, as in such a scenario the small grains should remain in the cavity while the millimeter sized grains are trapped at the edge. Given the presence of large structures at smaller radii, photoevaporation falls even shorter of explaining the origin of this second gap. Likewise, viscous flows of material to smaller radii seem unlikely, as this would raise new questions of why such a mechanism has preferentially cleared this region but not those at immediately larger or smaller radii. Hence we are left to consider the intriguing and seemingly likely possibility that this second gap is a direct result of the formation of planets and their ongoing sculpting of this region of the disk. 

\subsection{Companions Within the Disk}

Finally, there exists the possibility for the presence of multiple orbiting companions which may be responsible for the clearing of material, as well as sending material into orbit at altitude above the mid-plane (which may explain the origin of the mini-envelope that we have found to be a necessary component). Such companions may be on highly eccentric orbits, and at the inner parts of their orbits may travel close enough to perturb the sub-AU disk structures on periods consistent with the $\sim$10 yr timescale of the change in NIR emission. Furthermore, \cite{biller14} \& \cite{reggiani14} have reported the independent L' detection of planet-candidates respectively at separations of $\sim$0.11'' and 0.156"$\pm$0.032'' from the star. At a distance of 145 pc, these angular separations correspond to physical separations of $\sim$16 AU and 22.7$\pm$4.7 AU, placing them within the inner gap. Both teams present observations lacking corresponding detections of the L' hot-spots at shorter wavelengths, thus inferring their origin as sub-stellar. However, each team detects only one hot-spot, at differing radii, magnitude, and position-angle. Hence the validation of these planet-candidates presently remains open to skepticism. Incidentally, neither team detected L' counterpart emission from the compact 7 mm source in \cite{osorio14}. At 7 mm the disk is optically thin, raising the possibility of a chance coincidence with a background object. To be compelling, the L' candidates need to be recovered again, and also seen in other filters (such as 3.1 microns).  Nevertheless, the structures within the disk provide compelling evidence of planetary interactions and warrant the continued observation of the HD 169142 system as a candidate for the direct imaging of young extra-solar planets. 

\section{Summary and Conclusions}

Our assembly of the broadband SED from data sets covering a quarter of a century reveals variability of up to 45\% at NIR wavelengths over a maximum timescale of 10 years. Through modeling of the SED we confirm several scenarios where changes to the inner (sub-AU) structures may reproduce the two distinct states of the NIR spectrum. The simplest of these involves a shift of the outer edge of the inner disk by $\sim$0.05 AU inward toward the star. In this scenario we have kept the density of the mid-plane the same by adjusting the total mass of the inner disk by 6.3$\times 10^{-8}$M$_{\odot}$ to account for the decrease in volume. From the SpeX observations we derive a mass acrretion rate of \.{M} $\approx$ (1.5 - 2.7) x 10$^{-9}$ M$_{\odot}$ yr$^{-1}$, which accounts for as much as 43\% of the total mass change in the above scenario, and is suggestive that other processes (such as sculpting due to planets near the inner edge of the gap) may be active in addition to accretion onto the star. However, we find that the pre-2000 SED may be equally reproduced if the density of the inner disk is allowed to increase, allowing for the change in mass to be equivalent to that lost due to accretion. In any event, we do not observe any variable shadowing of the outer disk, requiring that changes must be occurring within the outer edge of the inner disk, or to an optically thin structure above/below the mid-plane. In order to reproduce the level of variability we have found that the latter structure is a necessary component to provide heating of the outer portions of the inner disk, and is similar to components present in the models and predictions in other studies on related objects (\cite{sitko08}, \cite{ellerbroek14}, \cite{vinkovic14}).

Through a combination of modeling the broadband SED and multi-band imagery we explore the structures of the outer disk and in particular the nature of the second gap detected in the H-band by \cite{quanz13} and \cite{momose13}, and at 7 mm by \cite{osorio14}. Our models suggest that this ring in the NIR is not significantly in part caused by shadows cast by inward disk structures. Instead, we find that the dark ring in the NIR may be reproduced by a 0.3$\times$ scaling of density or a 0.86$\times$ scaling of the height within the region. As a density feature is required to match the 7 mm observations, we find that a model fitting the images at both bands is possible with a density scaling $\geq 0.3\times$ and a height scaling of $\geq 0.86\times$, which we leave for future studies to explore with the hopeful acquisition of additional mm-continuum imagery. In any event, this structure is highly suggestive of planetary formation within the region. We find that the most constrained outer disk structure in the SED is a wall at $\sim$25 AU, whose emission resembles a nearly single temperature black-body with T=120K. While maintaining fits to the SED, we find that the H-band polarized surface brightness profile $\propto r^{-2.6}$ between 85-120 AU \citep{quanz13} may be reproduced by a radial density profile $\propto r^{-1}$ and a scale height exponent $\propto r^{1.3}$, while a steeper profile $\propto r^{-3}$ may be reproduced with a steeper fall off to the density $\propto r^{-2}$. Similarly, we find that a density profile $\propto r^{-2}$ in the 35-70 AU gap reproduces the observed surface brightness profile $\propto r^{-3}$ reported by \cite{momose13} for the region where r$<$50 AU.

Finally, we conclude that the protoplanetary disk in HD 169142 is home to a multitude of structures which may indicate the activity of planetary formation processes, and perhaps even that the effects of such processes may be visible over a single person's lifetime. At the very least, it offers a unique perspective on the timescales and mechanisms by which such disks become cleared of material, and hence when and how planets have the opportunity to form. In any event, HD 169142 is a highly structured and evolving system, of which continued observation and investigation may reveal further details, as well as answer important questions about the formation and evolution of young planetary systems. 

\section{Acknowledgments}

This work was supported by NASA ADAP grant NNX09AC73G, Hubble Space Telescope grant HST-GO-13032, the IR\&D program at The Aerospace Corporation, and the University of Cincinnati Honors Program. We would like to thank Dr. Sascha Quanz, Dr. Munetake Momose, and the Astronomical Society of the Pacific Conference Series for permission to reproduce the images from their original publications. We would also like to thank the anonymous journal referee whose comments greatly improved this article. 

\section{Appendix A: Model Parameters}

Parameters listed here were used in the 2014-01-31 release of the Whitney code in construction of our models of the circumstellar disk in HD 169142. Specific references are given where we have used a value within the bounds given in the referenced material, otherwise it may be assumed that the values were chosen upon for this study in seeking a best fit to the SED and images. The post-2000 model parameters were used as a basis for the pre-2000 models and the changes to each model are detailed below.

\begin{deluxetable}{ccc}
\tablenum{A1}
\tabletypesize{\footnotesize}
\tablecolumns{3} 
\tablewidth{0pc} 
\tablecaption{Model Parameters\label{partab}} 
\tablehead{ 
\colhead{Parameter}    & \colhead{Value} & \colhead{Reference}}
\startdata 
\cutinhead{Stellar Parameters}
Spectral Type & A7V & a\\
T$_{eff}$ & 7500 K & b\\
Radius (R$_{star}$) & 1.75 R$_{\odot}$ / 1.6 R$_{\odot}$ & c / a\\
Mass & 1.65 M$_{\odot}$ & a\\
Distance & 145 pc & d\\
\cutinhead{Post-2000 Model Parameters}
Disk inclination & 13$^{\circ}$& e\\
Disk mass (total) & 0.003 M$_{\odot}$& f\\
Fraction of mass in inner disk & $5\times 10^{-5}$& \nodata\\
Disk accretion rate & 2.1$\times 10^{-9}$ M$_{\odot}$& g\\
Inner disk grain file & www005.par & (1)\\
Inner disk scale height normalization & 0.002 R$_{star}$&\nodata\\
Inner disk R$_{in}$ - R$_{out}$& 0.23 - 0.27 AU&\nodata\\
Inner disk radial density $\propto R^{-A}$& A=1.0&\nodata\\
Inner disk scale height $\propto R^{B}$ & B=1.12&\nodata\\
Envelope grain file & carbonstardust.par & (3)\\
Envelope R$_{in}$ - R$_{out}$ & 0.23 - 0.27 AU&\nodata\\
Envelope density & 2.5$\times 10^{-15}$ g cm$^{-3}$&\nodata\\
Floor density (outside 0.27 AU) & $10^{-23}$ g cm$^{-3}$&\nodata\\
Envelope bipolar cavity opening angle & 15$^{\circ}$&\nodata\\
Bipolar cavity z-intercept & -0.13 AU&\nodata\\
Outer disk grain file & www003.par & (2)\\
Outer disk small grains file & draine\_opac\_new.dat & (4)\\
Outer disk \% mass in small grains & 4\%&\nodata\\
Outer disk scale height normalization & 0.00675 R$_{star}$&\nodata\\
Outer disk R$_{in}$ -  R$_{out}$ & 25 - 250 AU& h\\
Outer disk radial density $\propto R^{-A}$ & A=1.0& i\\
Outer disk scale height $\propto R^{B}$ & B=1.3&\nodata\\
Gap R$_{in}$ - R$_{out}$ & 35 - 70 AU& h , j\\
Gap radial density $\propto R^{-A}$ & A=2.0&\nodata\\ 
Gap density scaling & 0.3$\times$ & \nodata\\
Scale for radial exponential density cutoff & 120 AU&\nodata\\
Number of radial grid cells & 400&\nodata\\
Number of theta (polar) grid cells & 197 &\nodata\\
Number of phi (azimuthal) grid cells & 2 &\nodata\\\\
\cutinhead{Changes to Pre-2000 Model A}
Inner disk R$_{out}$& 0.285 AU&\nodata\\
Envelope R$_{out}$ & 0.32 AU&\nodata\\
Fraction of mass in inner disk & $7.1\times 10^{-5}$&\nodata\\
\cutinhead{Changes to Pre-2000 Model B}
Bipolar cavity z-intercept & -0.06 AU&\nodata\\
Envelope density & 5$\times 10^{-15}$ g cm$^{-3}$&\nodata\\
Fraction of mass in inner disk & $5.58\times 10^{-5}$&\nodata\\
\cutinhead{Changes to Pre-2000 Model C}
Envelope density & $10^{-14}$ g cm$^{-3}$&\nodata\\
Fraction of mass in inner disk & $5.67\times 10^{-5}$&\nodata\\
\enddata
\tablecomments{\textbf{(1)} This is model 3 from \cite{wood02}. \textbf{(2)} This is model 1 from \cite{wood02}.\textbf{(3)} This is the model from \cite{clayton11}, which uses the size distribution of Mathis, Rumpl, and Nordsieck (1977) and optical constants of burnt benzene from \cite{zubko96}. \textbf{(4)} This model was computed by Bruce Draine (Draine \& Li 2007), and is discussed in Wood et al. (2008).}
\tablerefs{\textbf{a)} \cite{blondel06} \textbf{b)} \cite{guimaraes06} \textbf{c)} value adopted in this study to match the van der Veen 1898 photometry \textbf{d)} \cite{sylvester96} \textbf{e)} \cite{raman06} \textbf{f)} \cite{meeus10} \textbf{g)} this study \textbf{h)} \cite{quanz13} \textbf{i)} \cite{momose13} \textbf{j)} \cite{osorio14}.}
\end{deluxetable}
\clearpage

\section{Appendix B: Photometry Table}

To aid in our construction of the SED, we have gathered the following data from their respective sources.

\begin{deluxetable}{cccccc}
\tabletypesize{\footnotesize}
\tablecolumns{6} 
\tablenum{B1}
\tablewidth{0pc} 
\tablecaption{Photometric Data \label{photab}} 
\tablehead{ 
\colhead{Data ID}    & \colhead{$\lambda$ [$\mu$m]} &  \colhead{Flux [W m$^{-2}$]} & \colhead{Error [W m$^{-2}$]} & \colhead{Date} & \colhead {Reference}}
\startdata
vanderveen & 0.45 & 1.25E-11 & 1.25E-13 & 1989 & a \\
vanderveen & 0.55 & 1.11E-11 & 1.11E-13 & 1989 & a \\
vanderveen & 0.68 & 8.05E-12 & 8.05E-14 & 1989 & a \\
vanderveen & 0.90 & 5.90E-12 & 5.90E-14 & 1989 & a \\ 
NICMOS & 1.12 & 5.567E-12 & 5.567E-14 & 2005-04-30 & b \\
Sylvester & 1.22 & 4.04E-12 & \nodata & 1992-06-12 & c\\
Hales & 1.24 & 4.04E-12 & 8.08E-14 &  2003-04-29 & d\\
Malfait & 1.24 & 4.07E-12 & \nodata & 1992 & e\\
2MASS & 1.25 & 4.60E-12 & 5.54E-14 & 1998-07-19 & f\\
NICMOS & 1.60 & 2.836E-12 & 2.836E-14 & 2005-04-30 & b \\  
Malfait & 1.63 & 3.04E-12 & \nodata & 1992 & e\\
2MASS & 1.65 & 3.24E-12 & 1.06E-13 & 1998-07-19 & f\\     
Sylvester & 1.65 & 3.06E-12 & \nodata & 1992-06-12 & c\\
2MASS & 2.17 & 2.52E-12 & 4.45E-14 & 1998-07-19 & f\\ 
Sylvester & 2.18 & 2.05E-12 & \nodata & 1992-06-12 & c\\
Malfait & 2.19 & 2.20E-12 & \nodata & 1992 & e\\
WISE & 3.35 & 9.05E-13 & 7.70E-14 & 2010-03-26 & g\\
Sylvester & 3.55 & 1.18E-12 & \nodata & 1992-06-12 & c\\
Sylvester & 3.76 & 1.00E-12 & \nodata & 1992-06-12 & c\\
Malfait & 3.77 & 1.12E-12 & \nodata & 1992 & e\\
WISE & 4.60 & 6.11E-13 & 2.90E-14 & 2010-03-26 & g\\ 
Sylvester & 4.77 & 7.40E-13 & \nodata & 1992-06-12 & c\\  
Malfait & 4.80 & 6.27E-13 & \nodata & 1992 & e\\    
Rayjay & 10.8 & 6.58E-13 & 6.583E-14 & 1999-05-04 & h\\
WISE & 11.56 & 6.28E-13 & 6.30E-15 & 2010-03-26 & g\\ 
IRAS & 12 & 7.38E-13 & 2.95E-14 & 1983 & i\\
AKARI IRC & 18 & 3.817E-12 & 3.899E-14 & 2006 & j\\
Rayjay & 18.2 & 1.296E-12 & 1.296E-13 & 1999-05-04 & h\\
WISE & 22.09 & 2.01E-12 & 1.20E-14 & 2010-03-26 & g\\
IRAS & 25 & 2.21E-12 & 1.11E-13 & 1983 & i\\
Harvey & 50 & 1.374E-12 & 2.46E-13 & 1995 & k\\ 
IRAS & 60 & 1.48E-12 & 1.33E-13 & 1983 & i\\
AKARI FIS & 65 &  1.128E-12 & 4.371E-15 & 2006 & l\\
AKARI FIS & 90 & 6.663E-13 & 4.100E-14 & 2006 & l\\
IRAS & 100 & 7.03E-13 & 7.73E-14 & 1983 & i\\
AKARI FIS & 140 & 2.854E-13 & 4.049E-14 & 2006 & l\\
AKARI FIS & 160 & 2.900E-13 & 7.969E-15 & 2006 & l\\
\enddata 
\vspace{-.8cm}
\tablerefs{\textbf{a)} \cite{vanderveen89} \textbf{b)} this study \textbf{c)} \cite{sylvester96} \textbf{d)} \cite{hales06} \textbf{e)} \cite{malfait98} \textbf{f)} 2MASS All-Sky Point Source Catalog \textbf{g)} WISE All-Sky Source Catalog \textbf{h)} \citet{jaya01} \textbf{i)} IRAS Point-Source Catalog \textbf{j)} AKARI/IRC Point Source Catalog \textbf{k)} \citet{harvey95} \textbf{l)} AKARI/FIS Bright Source Catalog}
\end{deluxetable}

\end{document}